\pdfoutput=1
\documentclass[12pt,notitlepage]{amsart}%
\usepackage{amssymb}
\usepackage{amsfonts}
\usepackage{graphicx}
\usepackage{amscd}
\usepackage{graphicx}
\usepackage{amsmath}
\usepackage{cite}
\usepackage{amsxtra}
\usepackage{footmisc}%
\newtheorem{theorem}{Theorem}
\theoremstyle{plain}

\newtheorem{lemma}[theorem]{Lemma}

\newtheorem{proposition}[theorem]{Proposition}
\newtheorem{remark}[theorem]{Remark}

\numberwithin{equation}{section}
\numberwithin{theorem}{section}

\ifx\pdfoutput\relax\let\pdfoutput=\undefined\fi
\newcount\msipdfoutput
\ifx\pdfoutput\undefined\else
\ifcase\pdfoutput\else
\ifx\paperwidth\undefined\else
\ifdim\paperheight=0pt\relax\else\pdfpageheight\paperheight\fi
\ifdim\paperwidth=0pt\relax\else\pdfpagewidth\paperwidth\fi
\fi\fi\fi
\begin{document}
\title[$p$-Adic Jackiw-Rebbi Model]{$p$-Adic Dirac Equations and the Jackiw-Rebbi Model}
\author[Z\'{u}\~{n}iga-Galindo]{W. A. Z\'{u}\~{n}iga-Galindo}
\address{University of Texas Rio Grande Valley\\
School of Mathematical \& Statistical Sciences\\
One West University Blvd\\
Brownsville, TX 78520, United States}
\email{wilson.zunigagalindo@utrgv.edu}

\begin{abstract}
We present a p-adic version of the Jackiw-Rebbi model, in which the real line
is replaced by the p-adic numbers and the Dirac Hamiltonian by a non-local
operator built from the twisted Vladimirov operator. We show that p-adic
quantum mechanics, although non-local, yields a physical picture that agrees
with the Archimedean one, realized through mechanisms with no Archimedean
counterpart. The agreement is specific: for a constant mass the p-adic model
is gapped, with the same gap; it has a chiral symmetry, so its spectrum is
symmetric about zero energy; a change of sign of the mass binds states at an
interface; and at low energies its spectrum cannot be distinguished from the
Archimedean continuous spectrum. The mechanisms are not: the p-adic free
Hamiltonian is diagonalized by compactly supported wavelets, so every
stationary state is strictly localized; momentum takes values on a geometric
ladder of powers of the prime; and a zero-energy interface state exists only
for masses commensurate with that ladder. The standard Dirac equation does not
admit space-localized solutions because it contradicts Einstein's causality.
The model is realizable at every finite resolution, as a continuous-time
quantum walk, and the topological protection of the Archimedean zero mode is lost.

\end{abstract}
\maketitle

\section{Introduction}

The hypothesis that spacetime at short (or very short) distances is discrete
is widely accepted, \cite{Capellmann,Amelino-Camelia,Bronstein,Rovelli,Volovich,Varadarajan}. The most natural way
to explain this hypothesis is through the Bronstein inequality, which asserts
that the uncertainty in any length measurement is greater than or equal to the
Planck length, \cite{Bronstein}. This inequality is typically interpreted as
saying that below the Planck length, there are no intervals, just points. From
a mathematical perspective, the intervals are just connected subsets. So, the
physical space at the Planck length can be modeled as a topological space in
which the non-trivial connected subsets are the points. This type of space is
well-known in mathematics as a completely (or totally) disconnected
topological space. Examples of such spaces are the Cantor sets, the field of
$p$-adic numbers $\mathbb{Q}_{p}$, or, more generally, $p$-adic Lie groups,
etc., \cite{V-V-Z,Dragovivh et al,Kozyrev,Zuniga-Textbook,Serre}. There are several physical theories that
assume space discreteness, among them quantum gravity and $p$-adic quantum
mechanics. The drawback of all these theories is that they are quite difficult
to test experimentally.

$p$-Adic quantum mechanics (QM) was introduced in the 1980s by Volovich and
Vladimirov, \cite{V-V-QM}. The literature on $p$-adic QM is pervasive, and
here we cite just a few works, \cite{Volovich,Varadarajan,V-V-Z,Dragovivh et al}, \cite{Varadarajan-2001,Vourdas,Anashin,Aniello et al,Zelenov,Urban,Zuniga-QM-2,Zuniga-AP,Zuniga-PhA,Zuniga-Mayes,Zuniga-Chacon,Zuniga-CTQW}.
Here, $p$-adic QM means QM (in the sense of Dirac-von Neumann) on
configuration spaces of type $\mathbb{Q}_{p}^{N}$, so the time is a real
variable. $p$-Adic QM is a model of standard QM under the hypothesis that the
physical space has a discrete nature. For about 45 years, it was unknown
whether $p$-adic QM was merely a mathematical toy model without physical
content. Recently, the author established that a certain type of $p$-adic
Schr\"{o}dinger equations describes a continuous version of quantum walks. In
practice, the discretization of such equations yields continuous-time quantum
walks, which contain, as a particular case, those on graphs, which are widely
used in quantum computing, \cite{Zuniga-QM-2}, see also \cite{Zuniga-Mayes,Zuniga-Chacon,Zuniga-CTQW,Mulkne-Blumen,Venegas-Andraca}.

In this paper, we introduce a $p$-adic version of the Jackiw-Rebbi model (JRM)
in one and two dimensions, \cite{Jackiw-Rebbi,Goldstone,Su et al,S-Q-Shen-Book,Kotetes}. The
Jackiw-Rebbi model is a seminal model in quantum field theory that predicts
the emergence of zero-energy bound states (zero modes) and fractional charge;
it was introduced in 1976 by R. Jackiw and C. Rebbi. The $p$-adic version
keeps the mechanism of the Archimedean one, in which a change of sign of the
mass binds states at an interface, and it reproduces the physical picture that
this mechanism produces; but it realizes that picture by means that have no
Archimedean counterpart, and these means are the point of this paper. The
underlying mathematical frameworks differ radically. The $p$-adic Dirac
Hamiltonian is an integral operator whose domain of integration is the whole
space. Thus, it is non-local, allowing long-range
interactions. In the Archimedean case, the Dirac Hamiltonian involves standard
derivatives, which are local operators. The $p$-adic Dirac Hamiltonian admits
localized wavefunctions, i.e., functions with compact support; this is
impossible for Archimedean Dirac Hamiltonians because it contradicts
Einstein's causality \cite{Thaller}.

We emphasize that no physical principle is violated here. The Archimedean
obstruction rests on the analyticity of the Fourier transform of a compactly
supported function, which is a consequence of the connectedness of the real
line; over $\mathbb{Q}_{p}$ there is no connectedness, and no notion of
analytic continuation of this type, so the argument has no $p$-adic analogue.
The $p$-adic model is not an Archimedean theory violating a bound; it is a
different theory, whose predictions, as we show, agree with the Archimedean
ones in the regime where the two can be compared.

The central physical contribution of this paper is Theorem
\ref{Thm_localization}: the free $p$-adic Dirac Hamiltonian is diagonalized by
the Kozyrev wavelets, \cite{Kozyrev}, which are compactly supported;
consequently it has pure point spectrum and a complete orthonormal family of
strictly localized stationary states, while the free Archimedean Dirac
operator has purely absolutely continuous spectrum and no stationary states at
all. The diagonalization by wavelets has two further consequences, which we
develop in Section \ref{Section_physics}. The momentum of the $p$-adic
particle takes its values on a geometric ladder of powers of $p$, so that the
energy levels are indistinguishable from the continuous spectrum of the
Archimedean model well below the mass gap and separate by a fixed factor well
above it; and at an interface where the mass changes sign, a zero-energy bound
state exists only when the mass is itself commensurate with that ladder. None
of these mechanisms has an Archimedean analogue; and yet the physical picture
that they produce is the familiar one: a gap for a constant mass, a spectrum
symmetric about zero energy, bound states at an interface where the mass
changes sign, and a low-energy spectrum indistinguishable from the continuous
spectrum of the Archimedean model.

Theorem \ref{Prop_CTQW} discretizes the $p$-adic Hamiltonian as an operator
acting on functions supported on a ball around the origin. The approximations
are matrices that generate continuous-time quantum walks on finite
hierarchical graphs, in the sense of \cite{Zuniga-CTQW}. This allows the
construction of the walks, and it is the sense in which the model prescribes
implementable dynamics at every finite level of resolution. The same analysis
shows what is lost: the topological protection of the Archimedean zero mode
has no $p$-adic counterpart, for reasons we make precise in Section
\ref{Section_invariants}.

The model admits two inequivalent implementations, and we study both. The
Archimedean Hamiltonian is $-iv\partial_{x}\sigma_{x}+m(x)v^{2}\sigma_{z}$,
and there are two ways of replacing the derivative in it by $\widetilde
{\boldsymbol{D}}_{x}$ (the twisted Vladimirov operator). The operator $\widetilde{\boldsymbol{D}}_{x}$ has a
real symbol, hence it is self-adjoint, so the correct dictionary with the
Archimedean theory identifies $\widetilde{\boldsymbol{D}}_{x}$ with the
momentum operator $-i\partial_{x}$, and not $\widetilde{\boldsymbol{D}}_{x}$
with $\partial_{x}$; the first implementation is therefore the self-adjoint
Hamiltonian $\boldsymbol{H}_{\mathrm{sa}}=v\widetilde{\boldsymbol{D}}%
_{x}\sigma_{x}+m(x)v^{2}\sigma_{z}$. The second comes from the substitution
$\partial_{x}\mapsto\widetilde{\boldsymbol{D}}_{x}$, which keeps the factor
$-i$ and gives the non-Hermitian operator $\boldsymbol{H}_{\mathrm{nh}%
}=-iv\widetilde{\boldsymbol{D}}_{x}\sigma_{x}+m(x)v^{2}\sigma_{z}$. It is the
second implementation that carries $E=0$ interface states, and whose interface
problem can be solved in closed form: such states exist precisely when the
masses satisfy the quantization condition mentioned above, and the
corresponding parameter values are exceptional points, where eigenvalues and
eigenfunctions coalesce.

The symmetry classification of the two models is sensitive to the arithmetic
of the prime: by a Gauss sum computation,
$\widetilde{\boldsymbol{D}}_{x}$ is odd under complex conjugation when
$p\equiv3$ $(\operatorname{mod}4)$ and even when $p\equiv1$
$(\operatorname{mod}4)$, and the antiunitary symmetries available to the model
change accordingly. Finally, in Section \ref{Section_invariants} we show that
the invariants of topological band theory are not available over
$\mathbb{Q}_{p}$: the $p$-adic Brillouin zone is totally disconnected, its
\v{C}ech cohomology vanishes in positive degrees and its odd $K$-theory is
trivial, so that there are no Chern or winding numbers to be computed. What
survives of the Jackiw-Rebbi phenomenon is its mechanism, not its topology.

We warn the reader that this paper is not about topological materials but
about the foundations of QM. Non-locality appears naturally in $p$-adic QM,
\cite{Zuniga-QM-2,Zuniga-AP,Zuniga-PhA,Zuniga-Mayes,Zuniga-Chacon}. The $p$-adic CTQWs of \cite{Zuniga-QM-2} already show that $p$-adic QM has a
physical meaning. The present paper adds a model in which a specific quantum
phenomenon is explained directly: the binding of zero-energy states at an
interface where the mass changes sign, together with the localization, the
momentum quantization and the mass selection rule that accompany it. On the other hand, in our view, the connection
between $p$-adic QM and topological materials opens a new avenue, as $p$-adic
analysis allows a unified treatment of discrete and continuous models.

\section{Basic facts on $p$-adic analysis}

In this section, we fix the notation and collect some basic results on
$p$-adic analysis that we will use throughout the article. For a detailed
exposition on $p$-adic analysis, the reader may consult \cite{V-V-Z}, \cite{Zuniga-Textbook}, \cite{Taibleson,Alberio et al}.

\subsection{The field of $p$-adic numbers}

From now on, we use $p$ to denote a fixed prime number. The field of $p-$adic
numbers $\mathbb{Q}_{p}$ is defined as the completion of the field of rational
numbers $\mathbb{Q}$ with respect to the $p-$adic norm $|\cdot|_{p}$, which is
defined as
\[
|x|_{p}=%
\begin{cases}
0 & \text{if }x=0\\
p^{-\gamma} & \text{if }x=p^{\gamma}\dfrac{a}{b},
\end{cases}
\]
where $a$ and $b$ are integers coprime with $p$. The integer $\gamma
=ord_{p}(x):=ord(x)$, with $ord(0):=+\infty$, is called the\textit{\ }$p-$adic
order of $x$. Any $p-$adic number $x\neq0$ has a unique expansion of the form
\[
x=p^{ord(x)}\sum_{j=0}^{\infty}x_{j}p^{j},
\]
where $x_{j}\in\{0,1,2,\dots,p-1\}$, $x_{0}\neq0$, and $\left\vert
x\right\vert _{p}=p^{-ord(x)}$. By using this expansion, we define the
fractional part $\{x\}_{p}$\textit{ of }$x\in\mathbb{Q}_{p}$ as the rational
number
\[
\{x\}_{p}=%
\begin{cases}
0 & \text{if }x=0\text{ or }ord(x)\geq0\\
p^{ord(x)}\sum_{j=0}^{-ord(x)-1}x_{j}p^{j} & \text{if }ord(x)<0.
\end{cases}
\]
In addition, any $x\in\mathbb{Q}_{p}\smallsetminus\left\{  0\right\}  $ can be
represented uniquely as $x=p^{ord(x)}v$, where $\left\vert v\right\vert
_{p}=1$. The elements of $\mathbb{Q}_{p}$ with norm $1$ (the units) form a
multiplicative group denoted as $\mathbb{Z}_{p}^{\times}$.

\subsection{Topology of $\mathbb{Q}_{p}$}

The field of $p$-adic numbers with the distance induced by $\left\vert
\cdot\right\vert _{p}$ is a complete ultrametric space. The ultrametric
property refers to the fact that $\left\vert x-y\right\vert _{p}\leq
\max\left\{  \left\vert x-z\right\vert _{p},\left\vert z-y\right\vert
_{p}\right\}  $ for any $x$, $y$, $z$ in $\mathbb{Q}_{p}$. This inequality
implies that if $\left\vert x\right\vert _{p}\neq\left\vert y\right\vert _{p}%
$, then $\left\vert x+y\right\vert _{p}=\max\left\{  \left\vert x\right\vert
_{p},\left\vert y\right\vert _{p}\right\}  $.

As a topological space $\left(  \mathbb{Q}_{p},\left\vert \cdot\right\vert
_{p}\right)  $ is totally disconnected, i.e., the only connected \ subsets of
$\mathbb{Q}_{p}$ are the empty set and the points. Furthermore, $\mathbb{Q}%
_{p}$\ is homeomorphic to a Cantor-like subset of the real line, see, e.g.,
\cite{V-V-Z}, \cite{Alberio et al}. $\mathbb{Q}_{p}$ is a paramount example of
a completely disconnected space, with a very rich mathematical structure. For
$r\in\mathbb{Z}$, denote by $B_{r}(a)=\{x\in\mathbb{Q}_{p};\left\vert
x-a\right\vert _{p}\leq p^{r}\}$ the ball of radius $p^{r}$ with center at
$a\in\mathbb{Q}_{p}$, and take $B_{r}(0):=B_{r}$. The balls are compact
subsets. Thus $\left(  \mathbb{Q}_{p},\left\vert \cdot\right\vert _{p}\right)
$ is a locally compact topological space.

\subsection{The Haar measure}

Since $(\mathbb{Q}_{p},+)$ is a locally compact topological group, there
exists a Haar measure $dx$, which is invariant under translations, i.e.,
$d(x+a)=dx$, \cite{Halmos}. If we normalize this measure by the condition
$\int_{\mathbb{Z}_{p}}dx=1$, then $dx$ is unique.

\subsection{The Bruhat-Schwartz space}

We will use $\Omega\left(  p^{-r}\left\vert x-a\right\vert _{p}\right)  $ to
denote the characteristic function of the ball $B_{r}(a)=a+p^{-r}%
\mathbb{Z}_{p}$, where
\[
\mathbb{Z}_{p}=\left\{  x\in\mathbb{Q}_{p};\left\vert x\right\vert _{p}%
\leq1\right\}
\]
is the $1$-dimensional unit ball. A complex-valued function $\varphi$ defined
on $\mathbb{Q}_{p}$ is called locally constant if for any $x\in\mathbb{Q}_{p}$
there exist an integer $l(x)\in\mathbb{Z}$ such that%
\begin{equation}
\varphi(x+x^{\prime})=\varphi(x)\text{ for any }x^{\prime}\in B_{l(x)}.
\label{local_constancy}%
\end{equation}
A function $\varphi:\mathbb{Q}_{p}\rightarrow\mathbb{C}$ is called a
Bruhat-Schwartz function\textit{ }(or a test function) if it is locally
constant with compact support. Any test function can be represented as a
linear combination, with complex coefficients, of characteristic functions of
balls. The $\mathbb{C}$-vector space of Bruhat-Schwartz functions is denoted
by $\mathcal{D}(\mathbb{Q}_{p})$.

\subsection{The Fourier transform}

Set $\chi_{p}(y)=\exp(2\pi i\{y\}_{p})$ for $y\in\mathbb{Q}_{p}$. The map
$\chi_{p}(\cdot)$ is an additive character on $\mathbb{Q}_{p}$, i.e., a
continuous map from $\left(  \mathbb{Q}_{p},+\right)  $ into $S$ (the unit
circle considered as multiplicative group) satisfying $\chi_{p}(x_{0}%
+x_{1})=\chi_{p}(x_{0})\chi_{p}(x_{1})$, $x_{0},x_{1}\in\mathbb{Q}_{p}$.\ 

The Fourier transform of $\varphi\in\mathcal{D}(\mathbb{Q}_{p})$ is defined
as
\[
\mathcal{F}\varphi(\xi)=%
{\displaystyle\int\limits_{\mathbb{Q}_{p}}}
\chi_{p}(\xi x)\varphi(x)dx\quad\text{for }\xi\in\mathbb{Q}_{p}.
\]
The Fourier transform is a linear isomorphism from $\mathcal{D}(\mathbb{Q}%
_{p})$ onto itself satisfying
\begin{equation}
(\mathcal{F}(\mathcal{F}\varphi))(\xi)=\varphi(-\xi). \label{Eq_FFT}%
\end{equation}
We will also use the notation $\mathcal{F}_{x\rightarrow\kappa}\varphi$ and
$\widehat{\varphi}$\ for the Fourier transform of $\varphi$.

\subsection{\label{Section_characters}Multiplicative characters}

From now on, we assume that $p$ is not equal to $2$. Any $x\in\mathbb{Q}%
_{p}^{\times}=\mathbb{Q}_{p}\smallsetminus\left\{  0\right\}  $ can be
uniquely represented as
\begin{equation}
x=p^{ord(x)}\mathrm{ac}(x)\text{, with\ }\mathrm{ac}(x)=x_{0}+x_{1}p+\ldots
\in\mathbb{Z}_{p}^{\times}. \label{polar_form}%
\end{equation}
We call $\mathrm{ac}(x)$ the angular component of $x$. We set
\[
\overline{\mathrm{ac}(x)}:=x_{0}\in\mathbb{F}_{p}^{\times}\text{,}%
\]
for the reduction $\operatorname{mod}$ $p$ of $\mathrm{ac}(x)$. We fix a
multiplicative character $\pi$ defined as%
\[%
\begin{array}
[c]{llll}%
\pi: & \mathbb{Q}_{p}^{\times} & \rightarrow & \left\{  \pm1\right\} \\
&  &  & \\
& x & \rightarrow & \left(  \frac{\overline{\mathrm{ac}(x)}}{p}\right)  ,
\end{array}
\]
where $\left(  \frac{\cdot}{p}\right)  $ is the Legendre symbol. Notice, that
by definition $\pi\left(  x\right)  =\pi\left(  \mathrm{ac}(x)\right)  $. For
the sake of completeness, we review the definition of $\left(  \frac{\cdot}%
{p}\right)  $: for $x_{0}\in\mathbb{F}_{p}^{\times}=\left\{  1,2,\ldots
,p-1\right\}  $,
\[
\left(  \frac{x_{0}}{p}\right)  =\left\{
\begin{array}
[c]{ll}%
1 & \text{if }z^{2}\equiv x_{0}\text{ }\operatorname{mod}p\text{ has solutions
in }\mathbb{F}_{p}^{\times}\\
& \\
-1 & \text{otherwise,}%
\end{array}
\right.
\]
where $a\equiv b$ $\operatorname{mod}p$ means that $a-b$ is divisible by $p$.
By definition $\left(  \frac{0}{p}\right)  =0$. Alternatively, $\left(
\frac{x_{0}}{p}\right)  =1$ means that $x_{0}$ is a square $\operatorname{mod}%
p$. We recall that for $p\neq2$, the cardinality of the squares in
$\mathbb{F}_{p}^{\times}$ (which is $\frac{p-1}{2}$) coincides with the
cardinality of non-squares in $\mathbb{F}_{p}^{\times}$.

We shall use Euler's criterion: for $p\neq2$ and $x_{0}\in\mathbb{F}%
_{p}^{\times}$,%
\[
\left(  \frac{x_{0}}{p}\right)  \equiv x_{0}^{\frac{p-1}{2}}\operatorname{mod}%
p\text{,}%
\]
see, e.g., \cite[Chapter 5]{Ireland-Rosen}. Taking $x_{0}=-1$,%
\begin{equation}
\pi\left(  -1\right)  =\left(  \frac{-1}{p}\right)  =\left(  -1\right)
^{\frac{p-1}{2}}=%
\begin{cases}
+1 & \text{if }p\equiv1\ (\operatorname{mod}4)\\
-1 & \text{if }p\equiv3\ (\operatorname{mod}4).
\end{cases}
\label{Euler_criterion}%
\end{equation}

We now define%
\[
\mathbb{F}_{p}^{+}=\left\{  j\in\mathbb{F}_{p}^{\times}\text{;}\left(
\frac{j}{p}\right)  =1\right\}  \text{, \ }\mathbb{F}_{p}^{-}=\left\{
j\in\mathbb{F}_{p}^{\times}\text{;}\left(  \frac{j}{p}\right)  =-1\right\}  ,
\]

\[
\mathbb{Q}_{p}^{+}=\left\{  x\in\mathbb{Q}_{p}^{\times}\text{; }\pi\left(
x\right)  =1\right\}  \text{, }\mathbb{Q}_{p}^{-}=\left\{  x\in\mathbb{Q}%
_{p}^{\times}\text{; }\pi\left(  x\right)  =-1\right\}  ,
\]
and%
\[
\mathbb{Z}_{p}^{+}=\mathbb{Q}_{p}^{+}\cap\mathbb{Z}_{p}^{\times}\text{,
\ \ }\mathbb{Z}_{p}^{-}=\mathbb{Q}_{p}^{-}\cap\mathbb{Z}_{p}^{\times}\text{.}%
\]

Intuitively, the set $\mathbb{Q}_{p}^{+}$ contains all the \textquotedblleft
positive\textquotedblright\ $p$-adic numbers, while \ $\mathbb{Q}_{p}^{-}$
contains the \textquotedblleft negative\textquotedblright\ ones. The field of
$p$-adic numbers is not an ordered field; for this reason, the notion of
positiveness introduced here is not an order relation in $\mathbb{Q}_{p}$.
Notice that
\[
\mathbb{Q}_{p}=\left\{  0\right\}  \text{ }%
{\displaystyle\bigsqcup}
\text{ }\mathbb{Q}_{p}^{+}\text{ }%
{\displaystyle\bigsqcup}
\text{ }\mathbb{Q}_{p}^{-}.
\]
Using (\ref{polar_form}),%
\[
\mathbb{Q}_{p}^{\times}=%
{\displaystyle\bigsqcup\limits_{r=-\infty}^{\infty}}
\text{ \ }p^{r}\mathbb{Z}_{p}^{\times}=%
{\displaystyle\bigsqcup\limits_{r=-\infty}^{\infty}}
\text{ }p^{r}\mathbb{Z}_{p}^{+}\text{ }%
{\displaystyle\bigsqcup}
{\displaystyle\bigsqcup\limits_{r=-\infty}^{\infty}}
\text{ \ }p^{r}\mathbb{Z}_{p}^{-},
\]

and
\[
\mathbb{Z}_{p}^{+}=%
{\displaystyle\bigsqcup\limits_{j\in\mathbb{F}_{p}^{+}}}
\left(  j+p\mathbb{Z}_{p}\right)  \text{, \ \ }\mathbb{Z}_{p}^{-}=%
{\displaystyle\bigsqcup\limits_{j\in\mathbb{F}_{p}^{-}}}
\left(  j+p\mathbb{Z}_{p}\right)  \text{.}%
\]

\subsection{\label{Section_Vladimirov}Twisted Vladimirov operator}

For $\alpha>0$, we define the operator%
\begin{equation}
\widetilde{\boldsymbol{D}}_{x}^{\alpha,\pi}\varphi\left(  x\right)
=\mathcal{F}_{\xi\rightarrow x}^{-1}\left(  \pi^{-1}\left(  \xi\right)  \text{
}\left\vert \xi\right\vert _{p}^{\alpha}\text{ }\mathcal{F}_{x\rightarrow\xi
}\varphi\right)  \text{, } \label{Taiblseon-Vladimirov_operator}%
\end{equation}
for $\varphi\in\mathcal{D}(\mathbb{Q}_{p})$. We call this operator the twisted
Vladimirov operator. This operator is non-local, more precisely,%
\[
\left(  \widetilde{\boldsymbol{D}}_{x}^{\alpha,\pi}\varphi\right)  \left(
x\right)  =%
{\displaystyle\int\limits_{\mathbb{Q}_{p}}}
\frac{\pi\left(  y\right)  \left\{  \varphi\left(  x-y\right)  -\varphi\left(
x\right)  \right\}  }{\Gamma_{p}\left(  -\alpha,\pi\right)  \left\vert
y\right\vert _{p}^{\alpha+1}}dy,
\]
where%
\[
\Gamma_{p}\left(  s,\pi\right)  =p^{s}%
{\displaystyle\int\limits_{\mathbb{Z}_{p}^{\times}}}
\pi\left(  t\right)  \chi_{p}\left(  p^{-1}t\right)  dt\text{,}%
\]
see \cite[Section 6.4.1]{Zuniga-LNM-2016}. We notice that%
\begin{align}
\Gamma_{p}\left(  s,\pi\right)   &  =p^{s}%
{\displaystyle\sum\limits_{j=1}^{p-1}}
\text{ }%
{\displaystyle\int\limits_{j+p\mathbb{Z}_{p}}}
\pi\left(  t\right)  \chi_{p}\left(  p^{-1}t\right)  dt=p^{s-1}%
{\displaystyle\sum\limits_{j=1}^{p-1}}
\left(  \frac{j}{p}\right)  e^{\frac{2\pi i}{p}j}\nonumber\\
&  =p^{s-1}\left\{
\begin{array}
[c]{ll}%
\sqrt{p} & \text{if }p\equiv1\operatorname{mod}4\\
i\sqrt{p} & \text{if }p\equiv3\operatorname{mod}4,
\end{array}
\right.  \label{Gauss_sum}%
\end{align}
see \cite[Chapter 6]{Ireland-Rosen}. We now set%
\[
\Phi_{rnj}\left(  x\right)  =p^{\frac{-r}{2}}\chi_{p}\left(  p^{-1}j\left(
p^{r}x-n\right)  \right)  \Omega\left(  \left\vert p^{r}x-n\right\vert
_{p}\right)  ,
\]
where $r\in\mathbb{Z}$, $j\in\mathbb{F}_{p}^{\times}=\left\{  1,2,\ldots
,p-1\right\}  $, $n\in\mathbb{Q}_{p}/\mathbb{Z}_{p}$. The Fourier transform of
$\Phi_{rnj}\left(  x\right)  $\ is given by%
\begin{equation}
\widehat{\Phi}_{rnj}\left(  \xi\right)  =p^{\frac{r}{2}}\chi_{p}\left(
p^{-r}n\xi\right)  \Omega\left(  \left\vert p^{-r}\xi+p^{-1}j\right\vert
_{p}\right)  . \label{FT_Psi}%
\end{equation}
Notice that $\Omega\left(  \left\vert p^{-r}\xi+p^{-1}j\right\vert
_{p}\right)  =1$ when $\xi\in p^{r-1}j+p^{r}\mathbb{Z}_{p}=$ \textrm{supp
}$\widehat{\Phi}_{rnj}\left(  \xi\right)  $, and that
\begin{align*}
\pi^{-1}  &  \mid_{p^{r-1}j+p^{r}\mathbb{Z}_{p}}=\pi^{-1}\left(  j\right)
\text{, }\\
\left\vert \xi\right\vert _{p}^{\alpha}  &  \mid_{p^{r-1}j+p^{r}\mathbb{Z}%
_{p}}=\left\vert p^{r-1}j+p^{r}\mathbb{Z}_{p}\right\vert _{p}=\left\vert
p^{r-1}j\right\vert _{p}^{\alpha}=p^{\left(  1-r\right)  \alpha}.
\end{align*}
Therefore from (\ref{Taiblseon-Vladimirov_operator}) and (\ref{FT_Psi}),
\[
\widetilde{\boldsymbol{D}}_{x}^{\alpha,\pi}\Phi_{rnj}\left(  x\right)
=\pi^{-1}\left(  j\right)  p^{\left(  1-r\right)  \alpha}\Phi_{rnj}\left(
x\right)  .
\]
We now set $\alpha=1$, $\widetilde{\boldsymbol{D}}_{x}^{1,\pi}:=\widetilde
{\boldsymbol{D}}_{x}$, and define%
\begin{equation}
\Theta_{rnj}\left(  x\right)  =\chi_{p}\left(  p^{-1}j\left(  p^{r}x-n\right)
\right)  \Omega\left(  \left\vert p^{r}x-n\right\vert _{p}\right)  .
\label{Theta_rnj}%
\end{equation}
Then%
\begin{equation}
\widetilde{\boldsymbol{D}}_{x}\Theta_{rnj}\left(  x\right)  =\pi^{-1}\left(
j\right)  p^{\left(  1-r\right)  }\Theta_{rnj}\left(  x\right)  .
\label{Eigenfunctions}%
\end{equation}
Thus $\Theta_{rnj}=p^{\frac{r}{2}}\Phi_{rnj}$ is the Kozyrev wavelet without
its normalization factor, and both functions are eigenfunctions of
$\widetilde{\boldsymbol{D}}_{x}$ with the same eigenvalue. We use
$\Phi_{rnj}$ where the orthonormality of the basis is needed, as in Theorems
\ref{Thm_localization} and \ref{Prop_CTQW}, and $\Theta_{rnj}$ in the
explicit computations of Sections \ref{Section_Hamiltonian}%
-\ref{Section_2D}, where $\Theta_{r0j}\left(  0\right)  =1$ makes the
matching of the solutions at the origin transparent; the normalization of
those solutions is imposed once, in Section \ref{Inreface}.

In \cite{Zuniga-PhA} we used the operator $\widetilde{\boldsymbol{D}}%
_{x}^{\alpha,1}$, that is, the trivial character. That operator is not
``chiral'': if $\boldsymbol{P}\varphi\left(  x\right)  =\varphi\left(
-x\right)  $, then $\widetilde{\boldsymbol{D}}^{\alpha,1}\boldsymbol{P}%
=\boldsymbol{P}\widetilde{\boldsymbol{D}}^{\alpha,1}$, cf. \cite[Lemma
4.1-(i)]{Zuniga-Mayes}, and for this reason we use $\widetilde{\boldsymbol{D}%
}^{\alpha,\pi}$ here. The $p$-adic Dirac equation built in \cite{Zuniga-PhA}
from $\widetilde{\boldsymbol{D}}_{x}^{\alpha,1}$ already shares many
properties with the standard one, among them particle-antiparticle pairs and a
charge conjugation symmetry, and it admits space-localized plane waves varying
in time, a phenomenon that does not occur in the Archimedean case, see, e.g.,
\cite[Section 1.8, Corollary 1.7]{Thaller}.

\section{\label{Section_dictionary}Self-adjointness, parity, and the role of
$p$ $\operatorname{mod}$ $4$}

Before writing a $p$-adic Dirac Hamiltonian, we must decide which Archimedean
operator $\widetilde{\boldsymbol{D}}_{x}$ replaces. The answer is dictated by
self-adjointness. Throughout the paper, we use units in which $\hbar=1$.

\subsection{$\widetilde{\boldsymbol{D}}_{x}$ is a momentum operator}

\begin{lemma}
\label{Lemma_selfadjoint}For $\alpha>0$, the operator $\widetilde
{\boldsymbol{D}}_{x}^{\alpha,\pi}$ defined in
(\ref{Taiblseon-Vladimirov_operator}) is self-adjoint on its natural domain
$\mathrm{Dom}=\{\varphi\in L^{2}(\mathbb{Q}_{p});\left\vert \xi\right\vert
_{p}^{\alpha}\widehat{\varphi}\in L^{2}(\mathbb{Q}_{p})\}$. Consequently
$-i\widetilde{\boldsymbol{D}}_{x}^{\alpha,\pi}$ is anti-self-adjoint.
\end{lemma}

\begin{proof}
The Fourier transform is a unitary operator of $L^{2}(\mathbb{Q}_{p})$, and
$\widetilde{\boldsymbol{D}}_{x}^{\alpha,\pi}$ is the operator of
multiplication, in the Fourier representation, by the function $\pi
^{-1}\left(  \xi\right)  \left\vert \xi\right\vert _{p}^{\alpha}$. Since $\pi$
takes the values $\pm1$, this function is real-valued, and multiplication by a
real-valued measurable function is self-adjoint on its maximal domain.
\end{proof}

The lemma is stated for every $\alpha>0$ because its proof does not use the
value of $\alpha$. In particular $\widetilde{\boldsymbol{D}}_{x}=\widetilde
{\boldsymbol{D}}_{x}^{1,\pi}$ is self-adjoint, and from here on we use only
this operator, following the convention of Section \ref{Section_Vladimirov}.

The Archimedean derivative $\partial_{x}$ is, in contrast, anti-self-adjoint:
its symbol $i\xi$ is purely imaginary. This is precisely why the Dirac
Hamiltonian carries the factor $-i$, the combination $-i\partial_{x}$ being
the (self-adjoint) momentum operator. Since $\widetilde{\boldsymbol{D}}_{x}$
is already self-adjoint, the correct dictionary is%
\[
-i\partial_{x}\longleftrightarrow\widetilde{\boldsymbol{D}}_{x}\text{,
\ \ that is, \ }\widehat{\boldsymbol{p}}:=\widetilde{\boldsymbol{D}}%
_{x}\text{,}%
\]
and not $\partial_{x}\leftrightarrow\widetilde{\boldsymbol{D}}_{x}$. Both
sides of the dictionary are self-adjoint operators with real, odd symbols (see
Lemma \ref{Lemma_Gauss} for the precise sense of odd). We use $\widehat
{\boldsymbol{p}}$ to build the self-adjoint model of Section
\ref{Section_Hamiltonian}; the non-Hermitian operator obtained by keeping the
factor $-i$ is studied in the same section, since, as we shall see, it is the
one that carries $E=0$ interface states.

\subsection{\label{Section_pmod4}Parity, complex conjugation and the prime}

We warn the reader that we use the bar over a symbol with two different
meanings. For $x\in\mathbb{Z}_{p}^{\times}$, $\overline{x}$ denotes the first
digit in the $p$-adic expansion of $x$, as in the notation $\overline
{\mathrm{ac}(x)}$ of Section \ref{Section_characters}; if $\varphi
:\mathbb{Q}_{p}\rightarrow\mathbb{C}$, $\overline{\varphi}$ denotes the
complex conjugate of $\varphi$.

We define the parity operator as $\boldsymbol{P}\varphi\left(  x\right)
=\varphi\left(  -x\right)  $, and the complex conjugation operator in the
$x$-representation as $\boldsymbol{K}\varphi\left(  x\right)  =\overline
{\varphi\left(  x\right)  }$, for $\varphi\in L^{2}\left(  \mathbb{Q}%
_{p}\right)  $. Notice that $\boldsymbol{P}^{2}=\boldsymbol{K}^{2}%
=\boldsymbol{I}$, and that $\left\Vert \boldsymbol{P}\right\Vert =\left\Vert
\boldsymbol{K}\right\Vert =1$.

\begin{lemma}
\label{Lemma_Gauss}(i) For $\varphi\in\mathrm{Dom}$,
\[
\boldsymbol{P}\widetilde{\boldsymbol{D}}_{x}\boldsymbol{P}=\pi\left(
-1\right)  \widetilde{\boldsymbol{D}}_{x}\text{, \ \ }\boldsymbol{K}%
\widetilde{\boldsymbol{D}}_{x}\boldsymbol{K}=\pi\left(  -1\right)
\widetilde{\boldsymbol{D}}_{x}\text{.}%
\]
(ii) The commutators $\left[  \widetilde{\boldsymbol{D}}_{x},\boldsymbol{P}%
\right]  =0$, $\left[  \widetilde{\boldsymbol{D}}_{x},\boldsymbol{K}\right]
=0$ if and only if $p\equiv1\ (\operatorname{mod}4)$.
\end{lemma}

\begin{proof}
(i) Since the space of test functions $\mathcal{D}(\mathbb{Q}_{p})$ is dense
in $\mathrm{Dom}\subset L^{2}\left(  \mathbb{Q}_{p}\right)  $, we may assume
without loss of generality that $\varphi\in\mathcal{D}(\mathbb{Q}_{p})$. Now,%
\begin{align*}
\widetilde{\boldsymbol{D}}_{x}\boldsymbol{P\varphi}\left(  x\right)   &
={\displaystyle\int\limits_{\mathbb{Q}_{p}}}\chi_{p}\left(  -\xi x\right)
\pi^{-1}\left(  \xi\right)  \text{ }\left\vert \xi\right\vert _{p}%
\mathcal{F}_{x\rightarrow\xi}(\boldsymbol{P}\varphi\left(  x\right)  )d\xi\\
&  ={\displaystyle\int\limits_{\mathbb{Q}_{p}}}\chi_{p}\left(  -\xi x\right)
\pi^{-1}\left(  \xi\right)  \text{ }\left\vert \xi\right\vert _{p}\left(  \mathcal{F}\varphi\right)  \left(  -\xi\right)  d\xi\\
&  =\pi^{-1}\left(  -1\right)  {\displaystyle\int\limits_{\mathbb{Q}_{p}}}%
\chi_{p}\left(  \xi x\right)  \pi^{-1}\left(  \xi\right)  \text{ }\left\vert
\xi\right\vert _{p}\left(  \mathcal{F}\varphi\right)  \left(
\xi\right)  d\xi=\pi^{-1}\left(  -1\right)  \boldsymbol{P}\left(
\widetilde{\boldsymbol{D}}_{x}\varphi\left(  x\right)  \right)  .
\end{align*}
On the other hand, using that $\mathcal{F}(\overline{\varphi})=\overline
{\mathcal{F}^{-1}(\varphi)}$,%
\begin{align*}
\widetilde{\boldsymbol{D}}_{x}\boldsymbol{K\varphi}\left(  x\right)   &
={\displaystyle\int\limits_{\mathbb{Q}_{p}}}\chi_{p}\left(  -\xi x\right)
\pi^{-1}\left(  \xi\right)  \text{ }\left\vert \xi\right\vert _{p}%
\mathcal{F}_{x\rightarrow\xi}(\boldsymbol{K}\varphi\left(  x\right)  )d\xi\\
&  ={\displaystyle\int\limits_{\mathbb{Q}_{p}}}\chi_{p}\left(  -\xi x\right)
\pi^{-1}\left(  \xi\right)  \text{ }\left\vert \xi\right\vert _{p}\left(  \overline{\mathcal{F}^{-1}\varphi
}\right)  \left(  \xi\right)  d\xi\\
&  =\overline{{\displaystyle\int\limits_{\mathbb{Q}_{p}}}\chi_{p}\left(  \xi
x\right)  \pi^{-1}\left(  \xi\right)  \text{ }\left\vert \xi\right\vert
_{p}\left(  \mathcal{F}^{-1}\varphi\right)  \left(  \xi\right)  d\xi
}\\
&  =\pi\left(  -1\right)  \overline{{\displaystyle\int\limits_{\mathbb{Q}_{p}%
}}\chi_{p}\left(  -\xi x\right)  \pi^{-1}\left(  \xi\right)  \text{
}\left\vert \xi\right\vert _{p}\left(  \mathcal{F}^{-1}%
\varphi\right)  \left(  -\xi\right)  d\xi}\\
&  =\pi\left(  -1\right)  \boldsymbol{K}\widetilde{\boldsymbol{D}}%
_{x}\boldsymbol{\varphi}\left(  x\right)  \text{,}%
\end{align*}
where in the last step we used that $\left(  \mathcal{F}^{-1}\varphi\right)
\left(  -\xi\right)  =\left(  \mathcal{F}\varphi\right)  \left(  \xi\right)
$. Since $\boldsymbol{P}^{2}=\boldsymbol{K}^{2}=\boldsymbol{I}$ and $\pi$
takes the values $\pm1$, the two identities of the statement follow.

(ii) The assertion about the commutators follows from Euler's criterion in the
form (\ref{Euler_criterion}). Indeed, for $p\equiv1\ (\operatorname{mod}4)$
the identities read $\boldsymbol{P}\widetilde{\boldsymbol{D}}_{x}%
=\widetilde{\boldsymbol{D}}_{x}\boldsymbol{P}$ and $\boldsymbol{K}%
\widetilde{\boldsymbol{D}}_{x}=\widetilde{\boldsymbol{D}}_{x}\boldsymbol{K}$.
If $p\not \equiv 1\ (\operatorname{mod}4)$, then $p\equiv
3\ (\operatorname{mod}4)$, and both operators anticommute with $\widetilde
{\boldsymbol{D}}_{x}$, which is not the zero operator, so that the commutators
do not vanish.
\end{proof}

\begin{lemma}
\label{Lemma_sign}For every $x\in\mathbb{Q}_{p}^{\times}$, $\pi\left(
-x\right)  =\pi\left(  -1\right)  \pi\left(  x\right)  $. Consequently, the
reflection $x\mapsto-x$ maps $\mathbb{Q}_{p}^{+}$ onto $\mathbb{Q}_{p}^{-}$ if
and only if $p\equiv3$ $(\operatorname{mod}4)$, while it maps each of
$\mathbb{Q}_{p}^{+}$, $\mathbb{Q}_{p}^{-}$ onto itself if and only if
$p\equiv1$ $(\operatorname{mod}4)$.
\end{lemma}

\begin{proof}
The identity $\pi\left(  -x\right)  =\pi\left(  -1\right)  \pi\left(
x\right)  $ holds because $\pi$ is multiplicative. If $p\equiv3$
$(\operatorname{mod}4)$, then $\pi\left(  -1\right)  =-1$ by
(\ref{Euler_criterion}), so $\pi\left(  -x\right)  =-\pi\left(  x\right)  $
and $x\in\mathbb{Q}_{p}^{\pm}$ implies $-x\in\mathbb{Q}_{p}^{\mp}$; if
$p\equiv1$ $(\operatorname{mod}4)$, then $\pi\left(  -1\right)  =1$ and
$\pi\left(  -x\right)  =\pi\left(  x\right)  $, so that each of $\mathbb{Q}%
_{p}^{+}$, $\mathbb{Q}_{p}^{-}$ is mapped onto itself. Since $p$ is odd, the
two cases are mutually exclusive and exhaust all the possibilities, so each
implication is an equivalence.
\end{proof}

\begin{remark}
\label{Remark_positivity}A different notion of positivity is used in
\cite[Section 6.2.5]{Zuniga-LNM-2016}: there $a\in\mathbb{Q}_{p}^{\times}$ is
declared positive when its first digit belongs to $\left\{  1,\ldots
,\frac{p-1}{2}\right\}  $, and negative otherwise. Since $p\neq2$, the digits
$z_{0}$ and $-z_{0}$ lie in different halves of $\mathbb{F}_{p}^{\times}$, so
with that convention the reflection $x\mapsto-x$ exchanges the positive and
the negative numbers for every odd prime, and no condition on $p$
$\operatorname{mod}$ $4$ is needed. The two conventions do not agree: for
$p=7$, for instance, the positive first digits are $\left\{  1,2,4\right\}  $
in the sense of Section \ref{Section_characters} and $\left\{  1,2,3\right\}
$ in the sense just recalled. We keep the convention of Section
\ref{Section_characters}, because the character $\pi$ that defines it is the
one that twists $\widetilde{\boldsymbol{D}}_{x}$, so that changing it would
change the operator itself, see Section \ref{Section_positivity}; with
that choice the reflection exchanges $\mathbb{Q}_{p}^{+}$ and $\mathbb{Q}%
_{p}^{-}$ exactly when $p\equiv3$ $(\operatorname{mod}4)$, by Lemma
\ref{Lemma_sign}. We note, in any case, that the arithmetic enters the
symmetry of Section \ref{Section_PT} through the operator as well as through
the mass profile: by Lemma \ref{Lemma_Gauss}, $\widetilde{\boldsymbol{D}}_{x}$
anticommutes with $\boldsymbol{P}$ precisely when $p\equiv3$
$(\operatorname{mod}4)$, whatever convention is adopted for the sign of a
$p$-adic number. The physical implications of this choice are discussed in
Section \ref{Section_positivity}.
\end{remark}

\subsection{\label{Section_PT}Parity, and the $\mathcal{PT}$-type symmetry in
the $p$-adic framework}

In Section 2.6 the quadratic character $\pi$ was used to split $\mathbb{Q}%
_{p}$ into the ``positive'' and ``negative'' parts $\mathbb{Q}_{p}^{+}$ and
$\mathbb{Q}_{p}^{-}$, and the mass profile of the $p$-adic Jackiw-Rebbi model,
given in Lemma \ref{Lemma_parity} below, is constant on each of them. The natural $p$-adic substitute for the Archimedean
reflection $x\mapsto-x$ is the parity operator $\boldsymbol{P}\varphi\left(
x\right)  =\varphi\left(  -x\right)  $ already used in Lemma \ref{Lemma_Gauss}%
. By Lemma \ref{Lemma_sign}, $\boldsymbol{P}$ exchanges $\mathbb{Q}_{p}^{+}$
and $\mathbb{Q}_{p}^{-}$ precisely when $p\equiv3$ $(\operatorname{mod}4)$.

In this section and in the rest of the paper we use the Pauli matrices%
\[
\sigma_{x}=\left[
\begin{array}
[c]{ll}%
0 & 1\\
1 & 0
\end{array}
\right]  \text{, \ }\sigma_{y}=\left[
\begin{array}
[c]{ll}%
0 & -i\\
i & 0
\end{array}
\right]  \text{, }\sigma_{z}=\left[
\begin{array}
[c]{ll}%
1 & 0\\
0 & -1
\end{array}
\right]  \text{,}%
\]
which in dimension one agree with the Dirac matrices. We write $\boldsymbol{1}%
$ for the identity, that is, for the $2\times2$ identity matrix and, when
acting on $L^{2}(\mathbb{Q}_{p})\times L^{2}(\mathbb{Q}_{p})$, for the
identity operator.

\begin{lemma}
\label{Lemma_parity}Let $p\equiv3$ $(\operatorname{mod}4)$, so that
$\pi\left(  -1\right)  =-1$ by (\ref{Euler_criterion}). Let $m(\cdot)$ be the
mass profile of the $p$-adic Jackiw-Rebbi model, that is, the function%
\[
m\left(  x\right)  =\left\{
\begin{array}
[c]{lll}%
-m_{1} &  & \text{if }x\in\mathbb{Q}_{p}^{-}\\
&  & \\
+m_{2} &  & \text{if }x\in\mathbb{Q}_{p}^{+},
\end{array}
\right.  =\left\{
\begin{array}
[c]{lll}%
\pi\left(  x\right)  m_{1} &  & \text{if }x\in\mathbb{Q}_{p}^{-}\\
&  & \\
\pi\left(  x\right)  m_{2} &  & \text{if }x\in\mathbb{Q}_{p}^{+},
\end{array}
\right.
\]
where $m_{1}$, $m_{2}$ are positive real numbers; it is locally constant on
$\mathbb{Q}_{p}^{+}$ and on $\mathbb{Q}_{p}^{-}$, and it is the $p$-adic
counterpart of the Archimedean profile of the Jackiw-Rebbi model, which equals
$-m_{1}$ for $x<0$ and $+m_{2}$ for $x>0$, see \cite[Section 2.2]%
{S-Q-Shen-Book}. This is the profile used in Section \ref{Section_Hamiltonian}. Then

\begin{enumerate}
\item[(i)] $\boldsymbol{P}\widetilde{\boldsymbol{D}}_{x}\boldsymbol{P}%
^{-1}=-\widetilde{\boldsymbol{D}}_{x}$;

\item[(ii)] $\boldsymbol{P}$ interchanges $\mathbb{Q}_{p}^{+}$ and
$\mathbb{Q}_{p}^{-}$, that is, it maps $\mathbb{Q}_{p}^{+}$ onto
$\mathbb{Q}_{p}^{-}$ and $\mathbb{Q}_{p}^{-}$ onto $\mathbb{Q}_{p}^{+}$, so
that $m\left(  -x\right)  =-m\left(  x\right)  $ when $m_{1}=m_{2}$;

\item[(iii)] consequently, if $m_{1}=m_{2}$, both Hamiltonians $\boldsymbol{H}%
_{\mathrm{sa}}=v\widetilde{\boldsymbol{D}}_{x}\sigma_{x}+m(x)v^{2}\sigma_{z}$
and $\boldsymbol{H}_{\mathrm{nh}}=-iv\widetilde{\boldsymbol{D}}_{x}\sigma
_{x}+m(x)v^{2}\sigma_{z}$ anticommute with $\boldsymbol{P}$,%
\[
\boldsymbol{PHP}^{-1}=-\boldsymbol{H}\text{,}%
\]
where $\boldsymbol{H}$ denotes $\boldsymbol{H}_{\mathrm{sa}}$ or
$\boldsymbol{H}_{\mathrm{nh}}$; furthermore, their spectra are symmetric with
respect to $E=0$.
\end{enumerate}
\end{lemma}

\begin{proof}
(i) is the first formula of Lemma \ref{Lemma_Gauss} together with $\pi\left(
-1\right)  =-1$. (ii) is Lemma \ref{Lemma_sign}; when the two values of the
mass have the same modulus it gives $m\left(  -x\right)  =-m\left(  x\right)
$. (iii) Combine (i) and (ii) in (\ref{Two_Hamiltonians}). The symmetry of the
spectrum then follows in the standard way: the relation $\boldsymbol{PHP}%
^{-1}=-\boldsymbol{H}$ says that $\boldsymbol{H}$ and $-\boldsymbol{H}$ are
unitarily equivalent, so that $\sigma\left(  \boldsymbol{H}\right)
=-\sigma\left(  \boldsymbol{H}\right)  $; at the level of eigenfunctions, if
$\boldsymbol{H}\Psi=E\Psi$, then $\boldsymbol{H}\left(  \boldsymbol{P}%
\Psi\right)  =-\boldsymbol{PH}\Psi=-E\left(  \boldsymbol{P}\Psi\right)  $, so
the levels occur in pairs $\pm E$. This is the usual argument for a chiral, or
sublattice, symmetry, see, e.g., \cite[Section 2.2]{S-Q-Shen-Book},
\cite{Kotetes}; it applies verbatim to $\boldsymbol{H}_{\mathrm{nh}}$, whose
eigenvalues may be complex. The symmetry classes to which these operators
belong are discussed in Section \ref{Section_invariants}.
\end{proof}

For $p\equiv1$ $(\operatorname{mod}4)$ the parity operator does not do this
job: there $-1$ is a square, so that $\pi\left(  -1\right)  =1$, and
$\boldsymbol{P}$ preserves $\mathbb{Q}_{p}^{+}$ and $\mathbb{Q}_{p}^{-}$
separately and commutes with $\boldsymbol{H}$ instead of anticommuting with
it. Its place is taken by the map $x\mapsto ux$, where $u$ is any unit with
$\pi\left(  u\right)  =-1$: this map is measure-preserving, it changes the
sign of the symbol of $\widetilde{\boldsymbol{D}}_{x}$, and it exchanges
$\mathbb{Q}_{p}^{+}$ and $\mathbb{Q}_{p}^{-}$, since $\pi\left(  ux\right)
=-\pi\left(  x\right)  $. Such a unit exists for every odd prime, because
exactly half of the elements of $\mathbb{F}_{p}^{\times}$ are non-squares; the
parity operator is the case $u=-1$, available precisely when $p\equiv3$
$(\operatorname{mod}4)$.

We can now say precisely what the symmetry usually called $\mathcal{PT}$ is in
this framework. In the Archimedean setting, a $\mathcal{PT}$-symmetric
Hamiltonian is one that commutes with the antiunitary operator obtained by
composing a spatial reflection with complex conjugation,
\cite{Bender-Boettcher}. The essential point is not the reflection but the
antiunitarity: an antiunitary symmetry forces the spectrum to be invariant
under complex conjugation.

\begin{lemma}
\label{Lemma_PT}Let $\boldsymbol{K}$ denote complex conjugation in the
$x$-representation, and set%
\[
\mathcal{T}=%
\begin{cases}
\boldsymbol{K} & \text{if }p\equiv3\ (\operatorname{mod}4)\\
\sigma_{z}\boldsymbol{K} & \text{if }p\equiv1\ (\operatorname{mod}4).
\end{cases}
\]
Then $\mathcal{T}$ is antiunitary, $\mathcal{T}^{2}=\boldsymbol{1}$, and
$\mathcal{T}\boldsymbol{H}_{\mathrm{nh}}\mathcal{T}^{-1}=\boldsymbol{H}%
_{\mathrm{nh}}$ for every real, bounded, locally constant mass. Consequently
the spectrum of $\boldsymbol{H}_{\mathrm{nh}}$ is invariant under complex
conjugation: its eigenvalues are real or occur in complex-conjugate pairs.
\end{lemma}

\begin{proof}
By Lemma \ref{Lemma_Gauss}, $\boldsymbol{K}\widetilde{\boldsymbol{D}}%
_{x}\boldsymbol{K}=\pi\left(  -1\right)  \widetilde{\boldsymbol{D}}_{x}$. If
$p\equiv3$ $(\operatorname{mod}4)$, then $\pi(-1)=-1$ and $\boldsymbol{K}%
\left(  -iv\widetilde{\boldsymbol{D}}_{x}\sigma_{x}\right)  \boldsymbol{K}%
=iv\left(  -\widetilde{\boldsymbol{D}}_{x}\right)  \sigma_{x}=-iv\widetilde
{\boldsymbol{D}}_{x}\sigma_{x}$, while the mass term is real and $\sigma_{z}$
has real entries, so it is unchanged. If $p\equiv1$ $(\operatorname{mod}4)$,
then $\boldsymbol{K}$ leaves $\widetilde{\boldsymbol{D}}_{x}$ invariant and
therefore changes the sign of the kinetic term; conjugating in addition by
$\sigma_{z}$ restores it, since $\sigma_{z}\sigma_{x}\sigma_{z}=-\sigma_{x}$.
The mass term is again unchanged, being real and commuting with $\sigma_{z}$. The statement about the spectrum
is the standard consequence of the existence of an antiunitary symmetry.
\end{proof}

We use the term \emph{$\mathcal{PT}$-type symmetry} for the antiunitary
symmetry $\mathcal{T}$ of Lemma \ref{Lemma_PT}, the role of the Archimedean
parity \emph{$\mathcal{P}$} being played by the operator $\boldsymbol{P}$ of
Lemma \ref{Lemma_parity}. Its spontaneous breaking, that is, the passage from
real eigenvalues to complex-conjugate pairs, occurs at the exceptional point
described in Section \ref{Section_EP}.

Lemma \ref{Lemma_Gauss} has a consequence that has no Archimedean counterpart:
the behaviour of the $p$-adic Dirac operator under time reversal, and
therefore its Altland-Zirnbauer symmetry class, is determined by an arithmetic
property of the prime. For $p\equiv3$ $(\operatorname{mod}4)$ the operator
$\widetilde{\boldsymbol{D}}_{x}$ is odd under $\boldsymbol{K}$, exactly like
the Archimedean momentum $-i\partial_{x}$; for $p\equiv1$ $(\operatorname{mod}%
4)$ it is even, and behaves like a real symmetric hopping term. We return to
this in Section \ref{Section_invariants}.

\section{\label{Section_Hamiltonian}1D $p$-Adic Jackiw-Rebbi model}

Our formulation of the JRM follows closely \cite[Section 2.2]{S-Q-Shen-Book};
the Pauli matrices $\sigma_{x}$, $\sigma_{y}$, $\sigma_{z}$ are those of
Section \ref{Section_PT}. We write $v$ for the effective velocity that
replaces the speed of light $c$ when the Dirac equation is applied to solids.

By Lemma \ref{Lemma_selfadjoint}, the substitution $\partial_{x}%
\rightarrow\widetilde{\boldsymbol{D}}_{x}$ in the Archimedean Hamiltonian
$-iv\partial_{x}\sigma_{x}+m(x)v^{2}\sigma_{z}$ produces an operator that is
not symmetric. We therefore consider the two 1D Dirac-type Hamiltonians%
\begin{equation}
\boldsymbol{H}_{\mathrm{sa}}(x)=v\widetilde{\boldsymbol{D}}_{x}\sigma
_{x}+m\left(  x\right)  v^{2}\sigma_{z}\text{, \ \ }\boldsymbol{H}%
_{\mathrm{nh}}(x)=-iv\widetilde{\boldsymbol{D}}_{x}\sigma_{x}+m\left(
x\right)  v^{2}\sigma_{z}\text{,} \label{Two_Hamiltonians}%
\end{equation}
the first self-adjoint, the second not. In both of them $m\left(  x\right)  $
is a locally constant function on $\mathbb{Q}_{p}^{+}$ ${\bigsqcup}$
$\mathbb{Q}_{p}^{-}$. So, there are two coverings,%
\begin{equation}
\mathbb{Q}_{p}^{-}=%
{\displaystyle\bigsqcup\limits_{i\in\mathbb{N}}}
\left(  x_{i}^{-}+p^{l_{i}^{-}}\mathbb{Z}_{p}\right)  \text{ and \ }%
\mathbb{Q}_{p}^{+}=%
{\displaystyle\bigsqcup\limits_{i\in\mathbb{N}}}
\left(  x_{i}^{+}+p^{l_{i}^{+}}\mathbb{Z}_{p}\right)  , \label{Disjoint union}%
\end{equation}
where ${\bigsqcup}$ means that the unions are disjoint, such that%
\[
m(x)\mid_{x_{i}^{-}+p^{l_{i}^{-}}\mathbb{Z}_{p}}=-m_{i}^{-}\text{, with }%
m_{i}^{-}>0\text{,}%
\]
and
\[
m(x)\mid_{x_{i}^{+}+p^{l_{i}^{+}}\mathbb{Z}_{p}}=m_{i}^{+}\text{, with }%
m_{i}^{+}>0\text{.}%
\]
This type of function includes, as a particular case, the profile of Lemma
\ref{Lemma_parity},%
\begin{equation}
m(x)=\left\{
\begin{array}
[c]{ll}%
-m_{1} & \text{if }x\in\mathbb{Q}_{p}^{-}\\
& \\
m_{2} & \text{if }x\in\mathbb{Q}_{p}^{+},
\end{array}
\right.  \label{function_m(x)}%
\end{equation}
where $m_{1}$,$m_{2}$ are positive real numbers.

\subsection{\label{Section_nh}The non-Hermitian model}

We analyze $\boldsymbol{H}_{\mathrm{nh}}$ first, because it is the one that
supports interface states at $E=0$; the self-adjoint model is studied in
Section \ref{Section_sa}. Throughout this subsection $\boldsymbol{H}%
(x):=\boldsymbol{H}_{\mathrm{nh}}(x)$,%
\[
\boldsymbol{H}(x)=-iv\widetilde{\boldsymbol{D}}_{x}\sigma_{x}+m\left(
x\right)  v^{2}\sigma_{z}.
\]
This operator is non-local, so long-range actions are allowed.

The eigenvalue problem for $\boldsymbol{H}(x)$ takes the form%
\begin{equation}
\left[
\begin{array}
[c]{lll}%
m(x)v^{2} &  & -iv\widetilde{\boldsymbol{D}}_{x}\\
&  & \\
-iv\widetilde{\boldsymbol{D}}_{x} &  & -m(x)v^{2}%
\end{array}
\right]  \left[
\begin{array}
[c]{l}%
\varphi_{1}\left(  x\right) \\
\\
\varphi_{2}\left(  x\right)
\end{array}
\right]  =E\left[
\begin{array}
[c]{l}%
\varphi_{1}\left(  x\right) \\
\\
\varphi_{2}\left(  x\right)
\end{array}
\right]  , \label{Equation}%
\end{equation}
where $E\in\mathbb{R}$, and $x\in\mathbb{Q}_{p}^{\times}$.

\textbf{Case} $x\in\mathbb{Q}_{p}^{-}$, $m(x)=-m_{1}$ for $x\in\left(
x_{1}+p^{l_{1}}\mathbb{Z}_{p}\right)  \subset\mathbb{Q}_{p}^{-}$.

For the sake of simplicity, we use $-m_{1}$ instead of $-m_{1}^{-}$. In this
case, (\ref{Equation}) take the form
\[
\left[
\begin{array}
[c]{lll}%
-m_{1}v^{2} &  & -iv\widetilde{\boldsymbol{D}}_{x}\\
&  & \\
-iv\widetilde{\boldsymbol{D}}_{x} &  & m_{1}v^{2}%
\end{array}
\right]  \left[
\begin{array}
[c]{l}%
\varphi_{1}\left(  x\right) \\
\\
\varphi_{2}\left(  x\right)
\end{array}
\right]  =E\left[
\begin{array}
[c]{l}%
\varphi_{1}\left(  x\right) \\
\\
\varphi_{2}\left(  x\right)
\end{array}
\right]  ,
\]
i.e.,%
\begin{equation}
\left[
\begin{array}
[c]{lll}%
-E-m_{1}v^{2} &  & -iv\widetilde{\boldsymbol{D}}_{x}\\
&  & \\
-iv\widetilde{\boldsymbol{D}}_{x} &  & -E+m_{1}v^{2}%
\end{array}
\right]  \left[
\begin{array}
[c]{l}%
\varphi_{1}\left(  x\right) \\
\\
\varphi_{2}\left(  x\right)
\end{array}
\right]  =\left[
\begin{array}
[c]{l}%
0\\
\\
0
\end{array}
\right]  . \label{Dirac_system_1}%
\end{equation}

We look for a solution of the form%
\begin{equation}
\left[
\begin{array}
[c]{l}%
\varphi_{1}\left(  x\right) \\
\\
\varphi_{2}\left(  x\right)
\end{array}
\right]  =\left[
\begin{array}
[c]{l}%
\varphi_{1}^{-}\\
\\
\varphi_{2}^{-}%
\end{array}
\right]  \Theta_{r_{-}n_{-}j_{-}}\left(  x\right)  , \label{Negative_sol}%
\end{equation}
where $x\in$ \textrm{supp} $\Theta_{r_{-}n_{-}j_{-}}\left(  x\right)
=p^{-r_{-}}n_{-}+p^{-r_{-}}\mathbb{Z}_{p}$, and $\pi^{-1}\left(  j_{-}\right)
=-1$, and
\[
\left(  x_{1}+p^{l_{1}}\mathbb{Z}_{p}\right)  \cap\left(  p^{-r_{-}}%
n_{-}+p^{-r_{-}}\mathbb{Z}_{p}\right)  \neq\emptyset.
\]
Taking $p^{-r_{-}}$ sufficiently large, we may assume that%
\begin{equation}
\left(  p^{-r_{-}}n_{-}+p^{-r_{-}}\mathbb{Z}_{p}\right)  \subset\left(
x_{1}+p^{l_{1}}\mathbb{Z}_{p}\right)  , \label{condition_1}%
\end{equation}
and thus%
\begin{equation}
r_{-}<-l_{1}. \label{condition_2}%
\end{equation}
By replacing (\ref{Negative_sol}) in (\ref{Dirac_system_1}), and using
(\ref{Eigenfunctions}),
\[
\left[
\begin{array}
[c]{lll}%
-E-m_{1}v^{2} &  & iv\lambda_{-}\\
&  & \\
iv\lambda_{-} &  & -E+m_{1}v^{2}%
\end{array}
\right]  \left[
\begin{array}
[c]{l}%
\varphi_{1}^{-}\\
\\
\varphi_{2}^{-}%
\end{array}
\right]  =\left[
\begin{array}
[c]{l}%
0\\
\\
0
\end{array}
\right]  ,
\]

where $\lambda_{-}=p^{1-r_{-}}$. This system has a non-trivial solution if
\[
\det\left[
\begin{array}
[c]{lll}%
-E-m_{1}v^{2} &  & iv\lambda_{-}\\
&  & \\
iv\lambda_{-} &  & -E+m_{1}v^{2}%
\end{array}
\right]  =E^{2}-m_{1}^{2}v^{4}+v^{2}\lambda_{-}^{2}=0,
\]
i.e., if $\lambda_{-}=\frac{\pm\sqrt{m_{1}^{2}v^{4}-E^{2}}}{v}$. Then
\[
\lambda_{-}=\frac{\sqrt{m_{1}^{2}v^{4}-E^{2}}}{v},\text{ with }m_{1}^{2}%
v^{4}-E^{2}\geq0.
\]
Notice that the case $m_{1}^{2}v^{4}-E^{2}<0$ is impossible. Now, since%
\[
\left(  -E-m_{1}v^{2}\right)  \varphi_{1}^{-}+iv\lambda_{-}\varphi_{2}^{-}=0,
\]%
\begin{equation}
\left[
\begin{array}
[c]{l}%
\varphi_{1}\left(  x\right) \\
\\
\varphi_{2}\left(  x\right)
\end{array}
\right]  =\left[
\begin{array}
[c]{l}%
\frac{iv\lambda_{-}}{E+m_{1}v^{2}}\\
\\
1
\end{array}
\right]  \varphi_{2}^{-}\Theta_{r_{-}n_{-}j_{-}}\left(  x\right)  ,
\label{Solution-minus}%
\end{equation}
for \ $x\in p^{-r_{-}}n_{-}+p^{-r_{-}}\mathbb{Z}_{p}$.

\textbf{Case} $x\in\mathbb{Q}_{p}^{+}$, $m\left(  x\right)  =m_{2}$ for
$x\in\left(  x_{2}+p^{l_{2}}\mathbb{Z}_{p}\right)  \subset\mathbb{Q}_{p}^{+}$.

For the sake of simplicity, we use $m_{2}$ instead of $m_{2}^{+}$. We look for
a solution of\ the form%
\[
\left[
\begin{array}
[c]{l}%
\varphi_{1}\left(  x\right) \\
\\
\varphi_{2}\left(  x\right)
\end{array}
\right]  =\left[
\begin{array}
[c]{l}%
\varphi+\\
\\
\varphi_{2}^{+}%
\end{array}
\right]  \Theta_{r_{+}n_{+}j_{+}}\left(  x\right)  ,
\]
where $x\in$\textrm{supp} $\Theta_{rnj}\left(  x\right)  =p^{-r_{+}}%
n_{+}+p^{-r_{+}}\mathbb{Z}_{p}$, $\pi^{-1}\left(  j_{+}\right)  =1$, and
\[
\left(  x_{2}+p^{l_{2}}\mathbb{Z}_{p}\right)  \cap\left(  p^{-r_{+}}%
n_{+}+p^{-r_{+}}\mathbb{Z}_{p}\right)  \neq\emptyset.
\]
Taking $p^{-r_{+}}$ sufficiently large, we may assume that
\begin{equation}
\left(  p^{-r_{+}}n_{+}+p^{-r_{+}}\mathbb{Z}_{p}\right)  \subset\left(
x_{2}+p^{l_{2}}\mathbb{Z}_{p}\right)  , \label{condition_3}%
\end{equation}
and thus%
\begin{equation}
r_{+}<-l_{2}. \label{condition_4}%
\end{equation}

In this case, the condition for a non-trivial solution is%

\[
\det\left[
\begin{array}
[c]{lll}%
-E+m_{2}v^{2} &  & -iv\lambda_{+}\\
&  & \\
-iv\lambda_{+} &  & -E-m_{2}v^{2}%
\end{array}
\right]  =E^{2}-m_{2}^{2}v^{4}+v^{2}\lambda_{+}^{2}=0,
\]
with $\lambda_{+}=p^{1-r_{+}}$, then
\[
\lambda_{+}=\frac{\sqrt{m_{2}^{2}v^{4}-E^{2}}}{v}.
\]
Since
\[
\left(  -E+m_{2}v^{2}\right)  \varphi_{1}^{+}-iv\lambda_{+}\varphi_{2}^{+}=0,
\]%
\begin{equation}
\left[
\begin{array}
[c]{l}%
\varphi_{1}\left(  x\right) \\
\\
\varphi_{2}\left(  x\right)
\end{array}
\right]  =\left[
\begin{array}
[c]{l}%
\frac{iv\lambda_{+}}{m_{2}v^{2}-E}\\
\\
1
\end{array}
\right]  \varphi_{2}^{+}\Theta_{r_{+}n_{+}j_{+}}\left(  x\right)  ,
\label{Solution-plus}%
\end{equation}
for \ $x\in p^{-r_{+}}n_{+}+p^{-r_{+}}\mathbb{Z}_{p}$.

\subsubsection{\label{Inreface}The wavefunction at the interface}

Take $x=0$. Assume that $0\in$ \textrm{supp} $\Theta_{r_{\pm}n_{\pm}j_{\pm}%
}\left(  x\right)  $, then the solution, (\ref{Solution-minus}) and
(\ref{Solution-plus}), takes two values at the origin, and thus these values
must coincide. Notice that $\Theta_{r_{\pm}n_{\pm}j_{\pm}}\left(  0\right)
=1$ means that $0\in\left(  p^{-r_{+}}n_{+}+p^{-r_{+}}\mathbb{Z}_{p}\right)
\cap\left(  p^{-r_{-}}n_{-}+p^{-r_{-}}\mathbb{Z}_{p}\right)  $. The condition
$\Theta_{r_{\pm}n_{\pm}j_{\pm}}\left(  0\right)  =1$ requires that $n_{\pm}%
=0$. Then
\[
\left[
\begin{array}
[c]{l}%
\varphi+\\
\\
\varphi_{2}^{+}%
\end{array}
\right]  \Theta_{r_{+}0j_{+}}\left(  0\right)  =\left[
\begin{array}
[c]{l}%
\varphi_{1}^{-}\\
\\
\varphi_{2}^{-}%
\end{array}
\right]  \Theta_{r_{-}0j_{-}}\left(  0\right)  ,
\]
i.e.,%
\[
\left[
\begin{array}
[c]{l}%
\frac{iv\lambda_{+}}{m_{2}v^{2}-E}\\
\\
1
\end{array}
\right]  \varphi_{2}^{+}=\left[
\begin{array}
[c]{l}%
\frac{iv\lambda_{-}}{E+m_{1}v^{2}}\\
\\
1
\end{array}
\right]  \varphi_{2}^{-},
\]
which implies that $\varphi_{2}^{+}=\varphi_{2}^{-}$ and
\[
\frac{iv\lambda_{+}}{m_{2}v^{2}-E}=\frac{iv\lambda_{-}}{E+m_{1}v^{2}}\text{,}%
\]
i.e.,
\begin{equation}
\frac{\sqrt{m_{2}^{2}v^{4}-E^{2}}}{m_{2}v^{2}-E}=\frac{\sqrt{m_{1}^{2}%
v^{4}-E^{2}}}{E+m_{1}v^{2}}. \label{condition_5}%
\end{equation}
Notice that $E=0$ satisfies this last condition, and consequently, the
solution (\ref{Solution-minus}) and (\ref{Solution-plus}) can be written as%
\[
\varphi_{m_{1,2},r_{\pm}0j_{\pm}}\left(  x\right)  =\left\{
\begin{array}
[c]{lll}%
\left[
\begin{array}
[c]{l}%
i\\
\\
1
\end{array}
\right]  \varphi_{m_{1,2}}^{+}\Theta_{r_{+}0j_{+}}\left(  x\right)  &
\text{if} & x\in\mathbb{Q}_{p}^{+}\\
&  & \\
\left[
\begin{array}
[c]{l}%
i\\
\\
1
\end{array}
\right]  \varphi_{m_{1,2}}^{+}\Theta_{r_{-}0j_{-}}\left(  x\right)  &
\text{if} & x\in\mathbb{Q}_{p}^{-}.
\end{array}
\right.
\]
where $\Theta_{r_{\pm}0j_{\pm}}\left(  x\right)  =\chi_{p}\left(
p^{-1+r_{\pm}}j_{\pm}x\right)  \Omega\left(  p^{-r_{\pm}}\left\vert
x\right\vert _{p}\right)  $.

For every pair $(m_{1},m_{2})$, there are $\left(  \frac{p-1}{2}\right)  ^{2}%
$\ functions of the form $\varphi_{m_{1,2},r_{\pm}0j_{\pm}}\left(  x\right)
$, corresponding to the $\frac{p-1}{2}$ possible values of $j_{+}\in
\mathbb{F}_{p}^{+}$ and the $\frac{p-1}{2}$ possible values of $j_{-}%
\in\mathbb{F}_{p}^{-}$, which are chosen independently. These functions are
not linearly independent: a combination $\sum c_{j_{+}j_{-}}\varphi
_{m_{1,2},r_{\pm}0j_{\pm}}$ vanishes precisely when $\sum_{j_{-}}c_{j_{+}%
j_{-}}=0$ for every $j_{+}$ and $\sum_{j_{+}}c_{j_{+}j_{-}}=0$ for every
$j_{-}$, so that the space they span, which is the space of zero-energy states
attached to the pair $(m_{1},m_{2})$, has dimension $\left(  p-1\right)
-1=p-2$. The number of chiral zero modes is therefore smaller than the number
of elementary solutions, and for $p=3$ it equals one, as in the Archimedean
model. Then, the wavefunction
at the interface is the superposition\ of the $\varphi_{m_{1,2}r_{\pm}0j_{\pm
}}\left(  x\right)  $:%
\begin{equation}
\Psi_{\text{inter}}\left(  x\right)  ={\sum\limits_{m_{1,2}r_{\pm}0j_{\pm}}%
}\varphi_{m_{1,2}r_{\pm}0j_{\pm}}\left(  x\right)  .
\label{wavefunction_interface}%
\end{equation}
The condition at the origin is satisfied by each $\varphi_{m_{1,2},r_{\pm
}0j_{\pm}}\left(  x\right)  $ separately, so it does not relate the constants
of different summands; the general zero-energy solution is the series
(\ref{wavefunction_interface}) with arbitrary square-summable coefficients. We
take all of them equal to one, which gives the unique state, up to a scalar,
that is invariant under the dilations $x\mapsto ux$ with $\pi\left(  u\right)
=1$, since $\Theta_{r_{\pm}0j_{\pm}}\left(  ux\right)  =\Theta_{r_{\pm}%
0j_{\pm}u}\left(  x\right)  $ and the squares act transitively on
$\mathbb{F}_{p}^{+}$ and on $\mathbb{F}_{p}^{-}$; this is the state that we
normalize below.

Here, it is important to say that (\ref{wavefunction_interface}) is a series.
This is a consequence of the fact that we are using countable coverings in the
definition of the function $m(x)$, see (\ref{Disjoint union}). On the other
hand, $\Psi_{\text{inter}}\left(  x\right)  \in L^{2}(\mathbb{Q}_{p})\times
L^{2}(\mathbb{Q}_{p})$. Given $\left(  f,g\right)  \in L^{2}(\mathbb{Q}%
_{p})\times L^{2}(\mathbb{Q}_{p})$, we take%
\[
\left\Vert \left(  f,g\right)  \right\Vert ^{2}=\left\Vert f\right\Vert
_{2}^{2}+\left\Vert g\right\Vert _{2}^{2}.
\]
Then%
\begin{gather*}
\left\Vert \varphi_{m_{1,2},r_{\pm}0j_{\pm}}\left(  x\right)  \right\Vert
^{2}=%
{\displaystyle\int\limits_{\mathbb{Q}_{p}^{+}}}
2\left\vert \varphi_{m_{1,2}}^{+}\right\vert ^{2}\Omega\left(  p^{-r_{+}%
}\left\vert x\right\vert _{p}\right)  dx+%
{\displaystyle\int\limits_{\mathbb{Q}_{p}^{-}}}
2\left\vert \varphi_{m_{1,2}}^{+}\right\vert ^{2}\Omega\left(  p^{-r_{-}%
}\left\vert x\right\vert _{p}\right)  dx\\
=2\left\vert \varphi_{m_{1,2}}^{+}\right\vert ^{2}\left\{
{\displaystyle\int\limits_{\mathbb{Q}_{p}^{+}}}
\Omega\left(  p^{-r_{+}}\left\vert x\right\vert _{p}\right)  dx+%
{\displaystyle\int\limits_{\mathbb{Q}_{p}^{-}}}
\Omega\left(  p^{-r_{-}}\left\vert x\right\vert _{p}\right)  dx\right\}  ,
\end{gather*}
and
\[
\left\Vert \Psi_{\text{inter}}\left(  x\right)  \right\Vert ^{2}=%
{\displaystyle\sum\limits_{m_{1,2},r_{\pm}0j_{\pm}}}
2\left\vert \varphi_{m_{1,2}}^{+}\right\vert ^{2}\left\{
{\displaystyle\int\limits_{\mathbb{Q}_{p}^{+}}}
\Omega\left(  p^{-r_{+}}\left\vert x\right\vert _{p}\right)  dx+%
{\displaystyle\int\limits_{\mathbb{Q}_{p}^{-}}}
\Omega\left(  p^{-r_{-}}\left\vert x\right\vert _{p}\right)  dx\right\}  .
\]
Now, by using the formula of Lemma \ref{Lemma_measure} below,%
\begin{equation}%
{\displaystyle\int\limits_{\mathbb{Q}_{p}^{\pm}}}
\Omega\left(  p^{-r_{\pm}}\left\vert x\right\vert _{p}\right)  dx=\frac{1}%
{2}p^{r_{\pm}},\label{Formula}%
\end{equation}

\[
\left\Vert \Psi_{\text{inter}}\left(  x\right)  \right\Vert ^{2}=%
{\displaystyle\sum\limits_{m_{1,2},r_{\pm}0j_{\pm}}}
\left\vert \varphi_{m_{1,2}}^{+}\right\vert ^{2}\left\{  p^{r_{+}}+p^{r_{-}%
}\right\}  .
\]
Now, a crucial observation is that the number of the $\varphi_{m_{1,2}}^{+}$
is finite and independent of $r_{\pm}$, by Lemma \ref{Lemma_count} below.
Therefore,%
\[
\left\Vert \Psi_{\text{inter}}\left(  x\right)  \right\Vert ^{2}=\left(
{\displaystyle\sum\limits_{m_{1,2}}}
\left\vert \varphi_{m_{1,2}}^{+}\right\vert ^{2}\right)  \left(
{\displaystyle\sum\limits_{j_{\pm}}}
1\right)  \left(
{\displaystyle\sum\limits_{r_{+}}}
p^{r_{+}}+%
{\displaystyle\sum\limits_{r_{-}}}
p^{r_{-}}\right)  <\infty,
\]
because $\sum_{r_{\pm}}p^{r_{\pm}}<\infty$ due conditions (\ref{condition_2}%
)-(\ref{condition_4}).

In conclusion,%
\[
\Psi_{\text{inter}}\left(  x\right)  =\frac{1}{\left\Vert \Psi_{\text{inter}%
}\left(  x\right)  \right\Vert }%
{\displaystyle\sum\limits_{m_{1,2}r_{\pm}0j_{\pm}}}
\varphi_{m_{1,2},r_{\pm}0j_{\pm}}\left(  x\right)
\]
is the normalized wavefunction at the interface. 

\bigskip

\begin{lemma}
\label{Lemma_measure}For every $r_{\pm}\in\mathbb{Z}$,%
\[%
{\displaystyle\int\limits_{\mathbb{Q}_{p}^{\pm}}}
\Omega\left(  p^{-r_{\pm}}\left\vert x\right\vert _{p}\right)  dx=\frac{1}%
{2}p^{r_{\pm}}\text{.}%
\]
\end{lemma}

\begin{proof}
We establish formula (\ref{Formula}) for $\mathbb{Q}_{p}^{+}$; the computation
for $\mathbb{Q}_{p}^{-}$ is identical.%
\begin{gather*}%
{\displaystyle\int\limits_{\mathbb{Q}_{p}^{+}}}
\Omega\left(  p^{-r_{+}}\left\vert x\right\vert _{p}\right)  dx=%
{\displaystyle\int\limits_{{\bigsqcup\nolimits_{r=-r_{+}}^{\infty}}%
p^{r}\mathbb{Z}_{p}^{+}}}
dx\\
=\text{ }%
{\displaystyle\sum\limits_{r=-r_{+}}^{\infty}}
\text{ }%
{\displaystyle\int\limits_{p^{r}\mathbb{Z}_{p}^{+}}}
dx=%
{\displaystyle\sum\limits_{r=-r_{+}}^{\infty}}
\text{ }%
{\displaystyle\sum\limits_{j\in\mathbb{F}_{p}^{+}}}
\text{ \ }%
{\displaystyle\int\limits_{p^{r}\left(  j+p\mathbb{Z}_{p}\right)  }}
dx.
\end{gather*}
Changing variables as $x=p^{r}y$, $dx=p^{-r}dy$, in the last integral,%
\begin{gather*}%
{\displaystyle\int\limits_{\mathbb{Q}_{p}^{+}}}
\Omega\left(  p^{-r_{+}}\left\vert x\right\vert _{p}\right)  dx=%
{\displaystyle\sum\limits_{r=-r_{+}}^{\infty}}
\text{ }%
{\displaystyle\sum\limits_{j\in\mathbb{F}_{p}^{+}}}
\text{ }p^{-r}\text{\ }%
{\displaystyle\int\limits_{j+p\mathbb{Z}_{p}}}
dx\\
=%
{\displaystyle\sum\limits_{r=-r_{+}}^{\infty}}
\text{ }%
{\displaystyle\sum\limits_{j\in\mathbb{F}_{p}^{+}}}
\text{ }p^{-r-1}=\left(  \frac{p-1}{2}\right)  p^{-1}%
{\displaystyle\sum\limits_{r=-r_{+}}^{\infty}}
\text{ }p^{-r}\text{\ }\\
=\left(  \frac{p-1}{2}\right)  p^{-1}\frac{p^{r_{+}}}{1-p^{-1}}=\frac{1}%
{2}p^{r_{+}}\text{.}%
\end{gather*}
\end{proof}

\begin{lemma}
\label{Lemma_count}The number of the $\varphi_{m_{1,2}}^{+}$ is finite and
independent of $r_{\pm}$.
\end{lemma}

\begin{proof}
We first notice that%
\[
p^{r}\mathbb{Z}_{p}/p^{r+1}\mathbb{Z}_{p}=\left\{  p^{r}j;j\in\mathbb{F}%
_{p}\right\}  ,
\]
and thus%
\begin{equation}
p^{r}\mathbb{Z}_{p}=%
{\displaystyle\bigsqcup\limits_{j\in\mathbb{F}_{p}^{+}}}
\left(  p^{r}j+p^{r+1}\mathbb{Z}_{p}\right)
{\displaystyle\bigsqcup}
{\displaystyle\bigsqcup\limits_{j\in\mathbb{F}_{p}^{-}}}
\left(  p^{r}j+p^{r+1}\mathbb{Z}_{p}\right)
{\displaystyle\bigsqcup}
p^{r+1}\mathbb{Z}_{p}.\label{Partition}%
\end{equation}
Then, there exists $j_{+}\in\mathbb{F}_{p}^{+}$ such that%
\[
m(x)\mid_{p^{r}j_{+}+p^{r+1}\mathbb{Z}_{p}}=m_{2}\left(  p^{r}j_{+}\right)  ,
\]
due to (\ref{condition_3}), and there exists $j_{-}\in\mathbb{F}_{p}^{-}$ such
that%
\[
m(x)\mid_{p^{r}j_{-}+p^{r+1}\mathbb{Z}_{p}}=-m_{1}\left(  p^{r}j_{-}\right)  .
\]
due to (\ref{condition_3}). Since the number of balls in partition
(\ref{Partition}) is independent on $r$, the number of pairs $-m_{1}\left(
p^{r}j_{-}\right)  $, $m_{2}\left(  p^{r}j_{+}\right)  $ satisfying
(\ref{condition_5}) is finite and independent of $r$.
\end{proof}

\subsubsection{\label{Section_EP}Mass quantization and exceptional points}

Two features of the construction above deserve to be stated explicitly, since
they have no analogue in the Archimedean model.

First, the $E=0$ solution is not available for arbitrary masses. The matching
condition (\ref{condition_5}) at $E=0$ requires%
\[
\lambda_{+}=m_{2}v\text{ \ and \ }\lambda_{-}=m_{1}v\text{,}%
\]
while $\lambda_{\pm}=p^{1-r_{\pm}}$ runs over the spectrum $p^{\mathbb{Z}}$ of
$\widetilde{\boldsymbol{D}}_{x}$, which is a discrete multiplicative ladder.
Hence the interface state exists if and only if%
\begin{equation}
m_{1}v\in p^{\mathbb{Z}}\text{ \ and \ }m_{2}v\in p^{\mathbb{Z}}\text{.}
\label{mass_quantization}%
\end{equation}
The $p$-adic geometry quantizes the admissible masses. This is a sharp
prediction of the model, and it is the $p$-adic counterpart of the fact that,
in the Archimedean case, the decay length $1/(mv)$ can take any positive
value, whereas here it must be a power of $p$.

Second, the point $\lambda=mv$ is an exceptional point, that is, a parameter
value at which two eigenvalues and their eigenvectors coalesce, so that the
operator ceases to be diagonalizable; this cannot happen for a self-adjoint
operator, whose eigenvectors stay orthogonal when two eigenvalues meet. The
analysis is local,
and $m$ denotes here the value of the mass on the ball under consideration,
that is, $m=m_{2}$ on $\mathbb{Q}_{p}^{+}$ and $m=m_{1}$ on $\mathbb{Q}%
_{p}^{-}$; no relation between $m_{1}$ and $m_{2}$ is assumed, and each side
has its own exceptional point, at $\lambda_{+}=m_{2}v$ and $\lambda_{-}%
=m_{1}v$ respectively. (The particular case $m_{1}=m_{2}$ is distinguished for
a different reason, namely the additional antisymmetry of Lemma
\ref{Lemma_parity}.) On a ball where $m(x)=m$ is constant and $\widetilde
{\boldsymbol{D}}_{x}\Theta=\lambda\Theta$, the operator $\boldsymbol{H}%
_{\mathrm{nh}}$ acts on the two-dimensional space spanned by $\Theta$ as%
\[
\boldsymbol{h}\left(  \lambda\right)  =\left[
\begin{array}
[c]{cc}%
mv^{2} & -iv\lambda\\
-iv\lambda & -mv^{2}%
\end{array}
\right]  \text{, \ \ }\det\left(  \boldsymbol{h}-E\right)  =E^{2}-m^{2}%
v^{4}+v^{2}\lambda^{2}\text{,}%
\]
so that $E=\pm\sqrt{m^{2}v^{4}-v^{2}\lambda^{2}}$. The two eigenvalues are
real for $v\lambda<mv^{2}$, form a complex-conjugate pair for $v\lambda
>mv^{2}$, and coalesce at $v\lambda=mv^{2}$, where%
\[
\boldsymbol{h}\left(  mv\right)  =mv^{2}\left[
\begin{array}
[c]{cc}%
1 & -i\\
-i & -1
\end{array}
\right]  \text{, \ \ }\boldsymbol{h}\left(  mv\right)  ^{2}=\boldsymbol{0}%
\text{.}%
\]
The block is nilpotent: both eigenvalues vanish and the two eigenvectors
coalesce into one, so that $\lambda=mv$ is an exceptional point of second
order, in the terminology of non-Hermitian quantum mechanics. It is also the
threshold at which the $\mathcal{PT}$-type symmetry of Lemma \ref{Lemma_PT}
is spontaneously broken, \cite{Bender-Boettcher,Heiss,Ashida et al}. The $E=0$ interface states constructed in Section \ref{Inreface} are
exactly the states sitting at this exceptional point.

Two consequences should be recorded. (i) Since $\lambda$ is unbounded above on
$L^{2}(\mathbb{Q}_{p})$, most of the spectrum of $\boldsymbol{H}_{\mathrm{nh}%
}$ is purely imaginary: the model is in the $\mathcal{PT}$-broken phase except
on the finitely many scales with $v\lambda\leq mv^{2}$. It describes a
dissipative, or amplifying, system rather than a conservative one, and its
time evolution $e^{-it\boldsymbol{H}_{\mathrm{nh}}}$ is not unitary. (ii) The
$E=0$ states are not protected in the sense of the Archimedean Jackiw-Rebbi
model: they exist only under the fine-tuning (\ref{mass_quantization}), and a
perturbation of the mass that violates it removes them. What replaces
protection here is the non-Hermitian notion of an exceptional point, which is
itself robust in a different sense: exceptional points are stable under
perturbations that preserve the $\mathcal{PT}$-type symmetry, and they are the
objects around which the topology of non-Hermitian systems is organized,
\cite{Bergholtz et al,Gong et al,Yao-Wang}.

\begin{figure}[th]
\centering
\includegraphics[width=\textwidth]{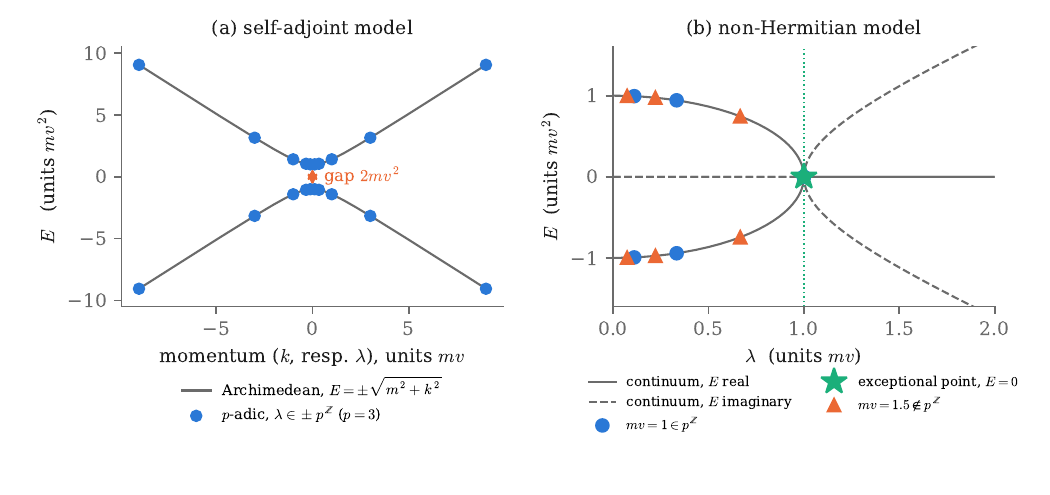}\caption{Energies of the
$p$-adic models compared with their Archimedean counterparts, in units
$m=v=1$, for $p=3$. (a) Self-adjoint model: the Archimedean dispersion
$E=\pm\sqrt{m^{2}v^{4}+v^{2}k^{2}}$ is a continuous curve defined for every
real momentum $k$, whereas the $p$-adic energies of Proposition \ref{Prop_gap}
are the values of the same function on the discrete multiplicative ladder
$\lambda\in\pm p^{\mathbb{Z}}$, the spectrum of $\widetilde{\boldsymbol{D}%
}_{x}$. The gap $2mv^{2}$ is the same in both cases, but the $p$-adic spectrum
is a pure point spectrum accumulating at $\pm mv^{2}$ from above and growing
geometrically. (b) Non-Hermitian model: the two energies $E=\pm\sqrt
{m^{2}v^{4}-v^{2}\lambda^{2}}$ are real below the exceptional point, imaginary
above it, and coalesce at $\lambda=mv$ (star). An $E=0$ interface state exists
only if some ladder point falls exactly on the star, i.e. only if $mv\in
p^{\mathbb{Z}}$ (circles); for a mass violating this condition (triangles) the
ladder straddles the exceptional point and no zero mode occurs. This is the
content of (\ref{mass_quantization}).}%
\label{Fig_dispersion}%
\end{figure}

\subsubsection{\label{Bulk_nh}The wavefunction at the bulk}

If the \textrm{supp} $\Theta_{r_{\pm}n_{\pm}j_{\pm}}\left(
x\right)  =p^{-r_{\pm}}n_{\pm}+p^{-r_{\pm}}\mathbb{Z}_{p}$ does not contain
the origin,
which means that distance from the origin to the ball $p^{-r_{\pm}}n_{\pm
}+p^{-r_{\pm}}\mathbb{Z}_{p}$ is positive, then the wavefunction outside of an
open compact subset containing the origin is a superposition of functions of
type:%
\begin{align*}
&  \left[
\begin{array}
[c]{l}%
\frac{iv\lambda_{+}}{m_{2}v^{2}-E}\\
\\
1
\end{array}
\right]  \varphi_{2}^{+}\Theta_{r_{+}n_{+}j_{+}}\left(  x\right)  ,\\
&  \text{ }\left[
\begin{array}
[c]{l}%
\frac{iv\lambda_{-}}{E+m_{1}v^{2}}\\
\\
1
\end{array}
\right]  \varphi_{2}^{-}\Theta_{r_{-}n_{-}j_{-}}\left(  x\right)  .
\end{align*}

\subsection{\label{Section_sa}The self-adjoint model}

We now turn to $\boldsymbol{H}_{\mathrm{sa}}=v\widetilde{\boldsymbol{D}}%
_{x}\sigma_{x}+m\left(  x\right)  v^{2}\sigma_{z}$, which is the Hamiltonian
obtained from the dictionary of Section \ref{Section_dictionary}. It is
self-adjoint whenever $m(\cdot)$ is a real, bounded, locally constant
function, and it has the chiral symmetry%
\begin{equation}
\Gamma=\sigma_{y}\text{, \ \ }\Gamma\boldsymbol{H}_{\mathrm{sa}}%
\Gamma=-\boldsymbol{H}_{\mathrm{sa}}\text{,} \label{chiral_symmetry}%
\end{equation}
because $\sigma_{y}$ anticommutes with $\sigma_{x}$ and with $\sigma_{z}$;
consequently its spectrum is symmetric with respect to $E=0$, for every mass profile.

\begin{proposition}
\label{Prop_gap}Let $m(x)=m$ be constant on a ball $B$ and let $\Theta
=\Theta_{rnj}$ be supported in $B$, with $\widetilde{\boldsymbol{D}}_{x}%
\Theta=\lambda\Theta$. Then the restriction of $\boldsymbol{H}_{\mathrm{sa}}$
to $\mathbb{C}^{2}\otimes\Theta$ has eigenvalues $E=\pm\sqrt{m^{2}v^{4}%
+v^{2}\lambda^{2}}$. In particular $\left\vert E\right\vert \geq\left\vert
m\right\vert v^{2}$, i.e., the model with constant mass is gapped, with gap
$2\left\vert m\right\vert v^{2}$.
\end{proposition}

\begin{proof}
On $\mathbb{C}^{2}\otimes\Theta$ the operator acts as the matrix
$v\lambda\sigma_{x}+mv^{2}\sigma_{z}$, and%
\[
\left(  v\lambda\sigma_{x}+mv^{2}\sigma_{z}\right)  ^{2}=\left(  v^{2}%
\lambda^{2}+m^{2}v^{4}\right)  \boldsymbol{1}\text{.}%
\]

\end{proof}

\begin{proposition}
\label{Prop_no_zero_mode}Let $m(\cdot)$ be locally constant. Then
$\boldsymbol{H}_{\mathrm{sa}}$ has no eigenfunction of the product form
$\psi=w\,\Theta_{rnj}$, $w\in\mathbb{C}^{2}\smallsetminus\{0\}$, with
eigenvalue $E=0$, provided $m$ does not vanish on the support of $\Theta
_{rnj}$.
\end{proposition}

\begin{proof}
Immediate from Proposition \ref{Prop_gap}: $E^{2}=m^{2}v^{4}+v^{2}\lambda
^{2}>0$.
\end{proof}

Thus the mechanism used in Section \ref{Inreface} is unavailable in the
self-adjoint model: there, the factor $-i$ turns $+v^{2}\lambda^{2}$ into
$-v^{2}\lambda^{2}$, which is what allows $E=0$. Proposition
\ref{Prop_no_zero_mode} rules out zero modes of the product form used in
Section \ref{Inreface}; it does not rule out interface states of a more
general form, and the following remark explains why that question is genuinely
harder in the $p$-adic setting than in the Archimedean one.

\begin{remark}
\label{Remark_origin}The mass profile (\ref{function_m(x)}) is locally
constant on $\mathbb{Q}_{p}^{+}$ and on $\mathbb{Q}_{p}^{-}$, but it is
\emph{not} locally constant at the origin: since $\pi$ takes both values on
every sphere $\left\vert x\right\vert _{p}=p^{k}$, every ball centred at $0$
meets both $\mathbb{Q}_{p}^{+}$ and $\mathbb{Q}_{p}^{-}$, however small it is.
In the Archimedean model the value of $m$ at the single point $x=0$ is
irrelevant, since a point has measure zero, and the wall separates two
half-lines. In the $p$-adic model the wall is not a point but a nested
sequence of spheres accumulating at $0$, and the two phases interleave at
every scale. Analytically, the consequence is that multiplication by
$m(\cdot)$ does not preserve the spaces of functions that are locally constant
of a given exponent, so the eigenvalue problem for $\boldsymbol{H}%
_{\mathrm{sa}}$ with the profile (\ref{function_m(x)}) does not decouple in
the basis $\{\Theta_{rnj}\}$, in contrast with the constant-mass case of
Proposition \ref{Prop_gap}. We regard the existence and the location of bound
states of $\boldsymbol{H}_{\mathrm{sa}}$ inside the gap as an open problem;
see Section \ref{Section_invariants}.
\end{remark}

Whatever the answer, those states cannot be pinned to $E=0$ by an index, as
they are in the Archimedean case: the chiral symmetry (\ref{chiral_symmetry})
forces the spectrum to be symmetric about $E=0$, but, as we show in Section
\ref{Section_invariants}, the winding number whose jump counts Jackiw-Rebbi
modes takes its values in a group that vanishes over a totally disconnected
configuration space. The $p$-adic self-adjoint model can therefore exhibit
interface states, but not topologically protected ones, and we regard this as
a structural feature rather than a defect of the model.

\subsubsection{The wavefunction at the bulk}

The same analysis applies to $\boldsymbol{H}_{\mathrm{sa}}$ away from the
origin. Let $B=p^{-r}n+p^{-r}\mathbb{Z}_{p}$ be a ball that does not contain
the origin, on which $m(\cdot)$ takes the constant value $m_{B}$, and let
$\Theta=\Theta_{rnj}$ be supported in $B$, so that $\widetilde{\boldsymbol{D}%
}_{x}\Theta=\lambda\Theta$ with $\lambda=\pi^{-1}\left(  j\right)  p^{1-r}$.
Then $\boldsymbol{H}_{\mathrm{sa}}$ leaves $\mathbb{C}^{2}\otimes\Theta$
invariant and acts on it as the matrix $v\lambda\sigma_{x}+m_{B}v^{2}%
\sigma_{z}$, whose eigenvalues are $E=\pm\sqrt{m_{B}^{2}v^{4}+v^{2}\lambda
^{2}}$ by Proposition \ref{Prop_gap}, with eigenvectors%
\[
\left[
\begin{array}
[c]{l}%
\frac{v\lambda}{E-m_{B}v^{2}}\\
\\
1
\end{array}
\right]  \varphi_{2}\Theta_{rnj}\left(  x\right)  \text{, \ \ }\varphi_{2}%
\in\mathbb{C}\text{.}%
\]
For the profile (\ref{function_m(x)}) this means $m_{B}=m_{2}$ when $B\subset
\mathbb{Q}_{p}^{+}$ and $m_{B}=-m_{1}$ when $B\subset\mathbb{Q}_{p}^{-}$. The
wavefunction outside an open compact subset containing the origin is a
superposition of functions of this type.

Two differences with the non-Hermitian case should be noted. First, the factor
$i$ is absent, so that the components of the spinor are real multiples of
$\varphi_{2}$ whenever $E$ is real. Second, by Proposition \ref{Prop_gap} we
have $\left\vert E\right\vert \geq\left\vert m_{B}\right\vert v^{2}>0$ on
every such ball, so that $\boldsymbol{H}_{\mathrm{sa}}$ has no solution at
$E=0$ and the construction of Section \ref{Inreface} has no self-adjoint
counterpart; the corresponding question at the origin is the open problem
recorded in Remark \ref{Remark_origin}.

\section{\label{Section_Compare}1D Jackiw-Rebbi model: Archimedean versus
$p$-Adic}

In this section, we compare the 1D $p$-adic Jackiw-Rebbi model against its
Archimedean counterpart. We follow closely the presentation given in
\cite[Sections 2.1-2.2]{S-Q-Shen-Book}. We assume that the function $m(x)$ has
the form (\ref{function_m(x)}).

\subsection{\textbf{Configuration space}}

The Archimedean Jackiw-Rebbi uses $\mathbb{R}$ as configuration space. In this
space, the intervals are connected subsets, which means that given two
different points from an interval, there is a point between them. The $p$-adic
version uses $\mathbb{Q}_{p}$, in this space space the balls (the intervals)
are not connected subsets. Thus, $\mathbb{R}$ is a continuous space, while
$\mathbb{Q}_{p}$ is discrete space.

\subsection{\textbf{Hamiltonians}}

The Jackiw-Rebbi 1D Hamiltonian for the Archimedean model is
\[
-iv\partial_{x}\sigma_{x}+m\left(  x\right)  v^{2}\sigma_{z},
\]
while the $p$-adic versions are the two operators (\ref{Two_Hamiltonians}).
There are two differences, and they should be kept apart. The first is
locality: $\partial_{x}$ is a local operator, while $\widetilde{\boldsymbol{D}%
}_{x}$ is non-local. Given $f:\mathbb{R}\rightarrow\mathbb{C},$ and $x_{0}%
\in\mathbb{R}$, the derivative $\partial_{x}f\left(  x_{0}\right)  $, depends
only of the behavior of the function $f(x)$ near to $x_{0}$. On the other
hand, the derivative $\widetilde{\text{ }\boldsymbol{D}}_{x}g\left(
y_{0}\right)  $ of a function $g:\mathbb{Q}_{p}\rightarrow\mathbb{C},$ at
$y_{0}\in\mathbb{Q}_{p}$, depends on all the points in the space
($\mathbb{Q}_{p}$). Then, the $p$-adic Hamiltonian allows long-range distance
interactions, while the Archimedean counterpart
does not allow it. The second difference is self-adjointness. The Archimedean
operator $-i\partial_{x}$ is self-adjoint because $\partial_{x}$ is not; the
$p$-adic operator $\widetilde{\boldsymbol{D}}_{x}$ is itself self-adjoint
(Lemma \ref{Lemma_selfadjoint}), so $\boldsymbol{H}_{\mathrm{sa}}$ is the
faithful counterpart of the Archimedean Hamiltonian and $\boldsymbol{H}%
_{\mathrm{nh}}$ is a genuinely different, non-Hermitian object. The parameter
$v$ and the Pauli matrices are the same in all three models.

\subsection{\textbf{The wavefunction at }$E=0$}

The two $p$-adic models must be distinguished here. By Proposition
\ref{Prop_gap} the self-adjoint model has no solution at $E=0$, since its
spectrum satisfies $\left\vert E\right\vert \geq\left\vert m\right\vert
v^{2}$ on every ball where the mass is constant; the comparison below
therefore concerns the non-Hermitian model $\boldsymbol{H}_{\mathrm{nh}}$ of
Section \ref{Section_nh}.

In the Archimedean case the wavefunction is%
\[
\Psi_{\mathbb{R}}\left(  x\right)  =\left\{
\begin{array}
[c]{lll}%
\sqrt{v\frac{m_{1}m_{2}}{m_{1}+m_{2}}}\left[
\begin{array}
[c]{l}%
i\\
1
\end{array}
\right]  e^{-m_{2}vx} & \text{if} & x>0\\
&  & \\
\sqrt{v\frac{m_{1}m_{2}}{m_{1}+m_{2}}}\left[
\begin{array}
[c]{l}%
i\\
1
\end{array}
\right]  e^{m_{1}vx} & \text{if} & x<0.
\end{array}
\right.
\]
The probability density $\left\vert \Psi_{\mathbb{R}}\left(  x\right)
\right\vert ^{2}$ is concentrated around $x=0$ ( or near to the interface).
The spatial distribution of $\Psi_{\mathbb{R}}\left(  x\right)  $ is
determined by the the characteristic scales $\lambda_{\pm}=\pm\frac{1}%
{m_{1,2}v}$. In practical terms, the wavefunction is \textquotedblleft
localized\textquotedblright\ on the interval $\left[  -\frac{1}{m_{1}v}%
,\frac{1}{m_{2}v}\right]  $. Furthermore, $\Psi_{\mathbb{R}}\left(  x\right)
$ vanish at $\pm\infty$. In the $p$-adic case, we display one of the
elementary solutions $\varphi_{m_{1,2},r_{\pm}0j_{\pm}}$ of Section
\ref{Inreface}, that is, the contribution of a single pair $\left(  j_{+}%
,j_{-}\right)  $, normalized to one; the wavefunction at the interface is the
superposition (\ref{wavefunction_interface}) of the $\left(  \frac{p-1}%
{2}\right)  ^{2}$ such solutions, and all of them have the same support and
the same characteristic scales, which are the features compared here. With
this understanding, the $p$-adic wavefunction takes the form%
\begin{equation}
\Psi\left(  x\right)  =\left\{
\begin{array}
[c]{lll}%
\sqrt{\frac{vm_{1}m_{2}}{p\left(  m_{1}+m_{2}\right)  }}\left[
\begin{array}
[c]{l}%
i\\
\\
1
\end{array}
\right]  \Theta_{r_{+}0j_{+}}\left(  x\right)  & \text{if} & x\in
\mathbb{Q}_{p}^{+}\\
&  & \\
\sqrt{\frac{vm_{1}m_{2}}{p\left(  m_{1}+m_{2}\right)  }}\left[
\begin{array}
[c]{l}%
i\\
\\
1
\end{array}
\right]  \Theta_{r_{-}0j_{-}}\left(  x\right)  & \text{if} & x\in
\mathbb{Q}_{p}^{-},
\end{array}
\right.  \label{Psi_Plus_minus}%
\end{equation}
where the support of $\Theta_{r_{+}0j_{+}}\left(  x\right)  $ is the ball
$B_{r_{+}}=\left\{  x\in\mathbb{Q}_{p};\left\vert x\right\vert _{p}\leq
p^{r_{+}}\right\}  $, with
\[
\lambda_{+}=p^{1-r_{+}}=m_{2}v\text{, i.e., }p^{r_{+}}=\frac{p}{m_{2}v},
\]
and the support of $\Theta_{r_{-}0j_{-}}\left(  x\right)  $ is the ball
$B_{r_{-}}=\left\{  x\in\mathbb{Q}_{p};\left\vert x\right\vert _{p}\leq
p^{r_{-}}\right\}  $, with
\[
\lambda_{-}=p^{1-r_{-}}=m_{1}v\text{, i.e., }p^{r_{-}}=\frac{p}{m_{1}v}.
\]

The wavefunction $\Psi\left(  x\right)  $ has compact support, it is really
localized. The support is the disjoint union of two balls with radii
$p^{r_{\pm}}$ controlled by the characteristic scales $p\frac{1}{m_{1,2}v}$.
The novelty is the occurrence of the factor $p$ (or $p^{-1}$) which depends on
the configuration space ($\mathbb{Q}_{p}$) chosen.

The full interface state is compactly supported as well. Each summand of
(\ref{wavefunction_interface}) is supported in a union of two balls centred at
the origin, of radii $p^{r_{\pm}}$, and the series%
\[%
{\displaystyle\sum\limits_{m_{1,2},r_{\pm}0j_{\pm}}}
\left(  p^{r_{+}}+p^{r_{-}}\right)
\]
converges, by the conditions (\ref{condition_1})-(\ref{condition_4}); in
particular these radii are bounded above, so that all the summands are
supported in one ball $B$, and $\Psi_{\text{inter}}$ is supported in $B$. For
the profile (\ref{function_m(x)}) the index $m_{1,2}$ takes a single value and
the support is exactly the union of the two balls of (\ref{Psi_Plus_minus}).

\subsection{\label{Section_role_p}The role of $p$}

The $p$-adic line $\mathbb{Q}_{p}$ is invariant under the transformations
$x\rightarrow ax+b$, with $a\in\mathbb{Q}_{p}^{\times}$ and $b\in
\mathbb{Q}_{p}$, and under them every non-zero $p$-adic number is equivalent
to one of the numbers%
\[
a_{0}+a_{1}p\text{, \ with }a_{0}\in\left\{  1,\ldots,p-1\right\}  \text{,
}a_{1}\in\left\{  0,1,\ldots,p-1\right\}  \text{:}%
\]
a transformation of type $x\rightarrow p^{L}x$ makes the norm equal to $1$,
and a translation removes the digits of order $p^{2}$ and higher. The smallest
distance between two different numbers of this form is $p^{-1}$. Therefore,
the constant $p$ has inverse length units. Then, the quantity $\frac{vm_{1}m_{2}%
}{p\left(  m_{1}+m_{2}\right)  }$, the square of the normalization constant in
(\ref{Psi_Plus_minus}), does not have units; it is just a number.

\subsection{Discussion}

The Archimedean and $p$-adic 1D Jackiw-Rebbi models both describe a Dirac
field coupled to a soliton field, through a spatially dependent mass term
whose sign changes. In the Archimedean case this sign flip gives rise to a
single zero-energy bound state at the interface, protected by chiral symmetry
and carrying fermion number $\pm\frac{1}{2}$, \cite{Jackiw-Rebbi}. The
comparison with the $p$-adic case is subtle; we summarize it as follows.

\begin{enumerate}
\item[(i)] \textbf{Localization.} Both $p$-adic models admit interface states
whose wavefunctions have compact support, which is impossible in the
Archimedean case; this is the sharpest difference between the two frameworks,
and it is a consequence of the ultrametric geometry rather than of the choice
of Hamiltonian.

\item[(ii)] \textbf{Zero energy.} The Archimedean zero mode sits at $E=0$ for
\emph{every} pair $m_{1}$, $m_{2}>0$. In the $p$-adic non-Hermitian model,
$E=0$ requires the quantization (\ref{mass_quantization}) and is realized at
an exceptional point. In the $p$-adic self-adjoint model, interface states
exist but sit at $E=\pm E_{\mathrm{mid}}$ with $E_{\mathrm{mid}}\neq0$ in
general (Section \ref{Section_sa}).

\item[(iii)] \textbf{Protection.} The Archimedean mode is protected by an
index: the number of chiral zero modes equals the jump of a winding number
across the wall. Over $\mathbb{Q}_{p}$ that index vanishes identically, for
the reasons given in Section \ref{Section_invariants}, and accordingly we make
no claim of topological protection.

\item[(iv)] \textbf{Fractional charge.} The fermion-number argument of Jackiw
and Rebbi uses the spectral symmetry of the Dirac Hamiltonian together with a
single isolated zero mode. In the $p$-adic self-adjoint model the spectral
symmetry holds, by (\ref{chiral_symmetry}), but the isolated zero mode does
not exist; we therefore do not claim charge fractionalization, and regard its
status in the $p$-adic framework as an open problem.
\end{enumerate}

In the $p$-adic model the Hamiltonian is a non-local operator, so long-range
interactions are allowed, and it admits localized quantum states, i.e.,
wavefunctions with compact support. These two features are absent in the
Archimedean model, and they are what makes the $p$-adic framework worth
developing, independently of the question of topological protection.

\section{\label{Section_2D}The 2D $p$-adic Jackiw-Rebbi model}

The 2D Hamiltonian for the Archimedean JRM has the form%
\[
m(x)v^{2}\sigma_{z}-iv\left(  \partial_{x}\sigma_{x}+\partial_{y}\sigma
_{y}\right)  ,
\]
see, e.g., \cite[Section 2.2.2]{S-Q-Shen-Book}. The assignment $\partial
_{x}\rightarrow\widetilde{\boldsymbol{D}}_{x}$ gives the right translation
from the 1D Archimedean JRM to the $p$-adic counterpart. We now extend this
analogy to the two-dimensional case. We recall that%
\begin{equation}
\partial=\partial_{x}+i\partial_{y} \label{Formula_D}%
\end{equation}
is the right derivative in the complex plane. Based on this fact, we propose%
\[
\widetilde{\boldsymbol{D}}=\widetilde{\boldsymbol{D}}_{x}+i\widetilde
{\boldsymbol{D}}_{y}%
\]
as $p$-adic analogue of (\ref{Formula_D}). We now propose the following
Hamiltonian for the 2D $p$-adic JRM:%
\[
\boldsymbol{H}(x,y)=m(x)v^{2}\sigma_{z}-iv\widetilde{\boldsymbol{D}}_{x}%
\sigma_{x}+v\widetilde{\boldsymbol{D}}_{y}\sigma_{y},
\]
which is the two-dimensional version of $\boldsymbol{H}_{\mathrm{nh}}$ in the
$x$-direction. Its self-adjoint counterpart, obtained from the dictionary of
Section \ref{Section_dictionary}, is%
\begin{equation}
\boldsymbol{H}_{\mathrm{sa}}(x,y)=m(x)v^{2}\sigma_{z}+v\left(  \widetilde
{\boldsymbol{D}}_{x}\sigma_{x}+\widetilde{\boldsymbol{D}}_{y}\sigma
_{y}\right)  \text{,} \label{H_2D_sa}%
\end{equation}
and the computation below goes through for it verbatim, with $E=v\pi
^{-1}(s)p^{1-l}$ replaced by the corresponding expression for the self-adjoint
model. We keep the non-Hermitian form in the main computation for continuity
with Section \ref{Section_Hamiltonian}. where $\sigma_{x}$, $\sigma_{y}$,
$\sigma_{z}$ are the $2\times2$ Pauli matrices and $m(x)$ is defined in
(\ref{function_m(x)}). We look for a wavefunction $\Psi^{\left(  2D\right)
}\left(  x,y\right)  $ satisfying%
\begin{equation}
\boldsymbol{H}(x,y)\Psi^{\left(  2D\right)  }\left(  x,y\right)
=E\Psi^{\left(  2D\right)  }\left(  x,y\right)  \text{.}
\label{Haamiltonian-2D}%
\end{equation}
Following, the Archimedean case, we propose a solution of the form%
\begin{equation}
\Psi^{\left(  2D\right)  }\left(  x,y\right)  =\Psi_{\pm}\left(  x\right)
\Theta_{lms}\left(  y\right)  , \label{Solution-2D}%
\end{equation}
where $\Psi_{\pm}\left(  x\right)  $ is the solution of the 1D model, see
(\ref{Psi_Plus_minus}), and $\Theta_{lms}\left(  x\right)  $ as in
(\ref{Theta_rnj}). Then, replacing (\ref{Solution-2D}) in
(\ref{Haamiltonian-2D}),%
\begin{gather*}
\boldsymbol{H}(x,y)\Psi_{\pm}\left(  x\right)  \Theta_{lms}\left(  y\right) \\
=\sqrt{\frac{vm_{1}m_{2}}{p\left(  m_{1}+m_{2}\right)  }}\Theta_{r_{\pm}0j_{\pm}}\left(
x\right)  v\left(  \widetilde{\boldsymbol{D}}_{y}\Theta_{lms}\left(  y\right)
\right)  \sigma_{y}\left[
\begin{array}
[c]{l}%
i\\
\\
1
\end{array}
\right] \\
=v\left(  \pi^{-1}\left(  s\right)  p^{1-l}\right)  \Psi^{\left(  2D\right)
}\left(  x,y\right)  .
\end{gather*}
Therefore,%
\[
E=v\left(  \pi^{-1}\left(  s\right)  p^{1-l}\right)  \text{, with }\pi
^{-1}\left(  s\right)  \in\left\{  \pm1\right\}  .
\]

We now fix $s_{+}\in\mathbb{F}_{p}^{+}$ (i.e., $\pi^{-1}\left(  s_{+}\right)
=1$), and $s_{-}\in\mathbb{F}_{p}^{-}$ (i.e., $\pi^{-1}\left(  s_{-}\right)
=-1$), to get%
\begin{equation}
\Psi_{\pm,+}^{\left(  2D\right)  }\left(  x,y\right)  =\left\{
\begin{array}
[c]{l}%
\sqrt{\frac{vm_{1}m_{2}}{p\left(  m_{1}+m_{2}\right)  }}\left[
\begin{array}
[c]{l}%
i\\
\\
1
\end{array}
\right]  \Theta_{r_{\pm}0j_{\pm}}\left(  x\right)  \Theta_{lms_{+}}(y),\\
\\
E=vp^{1-l},
\end{array}
\right.  \label{Psi_2D_1}%
\end{equation}%
\begin{equation}
\Psi_{\pm,-}^{\left(  2D\right)  }\left(  x,y\right)  =\left\{
\begin{array}
[c]{l}%
\sqrt{\frac{vm_{1}m_{2}}{p\left(  m_{1}+m_{2}\right)  }}\left[
\begin{array}
[c]{l}%
i\\
\\
1
\end{array}
\right]  \Theta_{r_{\pm}0j_{\pm}}\left(  x\right)  \Theta_{lms_{-}}(y),\\
\\
E=-vp^{1-l},
\end{array}
\right.  \label{Psi_2D_2}%
\end{equation}
$x,y\in\mathbb{Q}_{p}$.

\subsection{Discussion}

We review very quickly the 2D Archimedean JRM following \cite[Section
2.2.2]{S-Q-Shen-Book}. Again $m(x)=-m_{1}$ if $x>0$, otherwise $m(x)=m_{2}$,
with \ $m_{1}$, $m_{2}>0$. There are two wavefunctions distributed around the
interface: the first one%
\begin{equation}
\Psi_{\mathbb{R}}^{+}\left(  x,k_{y}\right)  =\sqrt{v\frac{m_{1}m_{2}}%
{m_{1+}m_{2}}}\left[
\begin{array}
[c]{l}%
i\\
0\\
0\\
1
\end{array}
\right]  e^{-\left\vert m(x)vx\right\vert +ik_{y}y}, \label{Psi_2D_3}%
\end{equation}
with dispersion $E_{+}=vk_{y}$, and the second
\begin{equation}
\Psi_{\mathbb{R}}^{-}\left(  x,k_{y}\right)  =\sqrt{v\frac{m_{1}m_{2}}%
{m_{1+}m_{2}}}\left[
\begin{array}
[c]{l}%
0\\
i\\
1\\
0
\end{array}
\right]  e^{-\left\vert m(x)vx\right\vert +ik_{y}y}, \label{Psi_2D_4}%
\end{equation}
with dispersion $E_{-}=-vk_{y}$. Each state carries a current along the
interface, but the electrons with different spins move in opposite directions.
The current density decays exponentially away from the interface. By comparing
(\ref{Psi_2D_1})-(\ref{Psi_2D_2}) with (\ref{Psi_2D_3})-(\ref{Psi_2D_4}), we
argue that the $p$-adic 2D\ version of the JRM provides the same physical
picture as the Archimedean one. In the $p$-adic 2D model, the domain wall is
no longer a point but an entire line
\begin{equation}
\left\{  0\right\}  \times\mathbb{Q}_{p}, \label{Line}%
\end{equation}
where the mass changes sign. The Jackiw-Rebbi (JR) states (\ref{Psi_2D_1}%
)-(\ref{Psi_2D_2}) are localized along the edge (\ref{Line}) that separates
two 2D materials. These states are the analogue of the chiral edge modes of
the Archimedean model, that is, of states associated with the boundary;
whether they propagate along it, in any sense that survives the passage to an
ultrametric space, is discussed in the remark at the end of this section. The
description the JR states is local, and it is based on the eigenfunctions of
the operator $\widetilde{\boldsymbol{D}}$. Notice that the wave numbers
$k_{y}=p^{1-l}$, $l\in\mathbb{Z}$, depend on the space scale ($p^{-1}$) of the
configuration space $\mathbb{Q}_{p}$.

One statement of the Archimedean picture does not transfer without further
work, and we state it as an open question rather than as a result. In the
Archimedean case, the two edge branches $E_{\pm}=\pm vk_{y}$ have opposite
group velocities $dE_{\pm}/dk_{y}=\pm v$, and this is what is meant by
counter-propagating chiral channels carrying opposite currents. Over
$\mathbb{Q}_{p}$ there is no group velocity: the dual variable $\xi$ is not a
real number, the dispersion $E=v\pi^{-1}(s)p^{1-l}$ is defined on the discrete
ladder $p^{\mathbb{Z}}$ and cannot be differentiated, and $\mathbb{Q}_{p}$
carries no orientation, so ``opposite directions'' has no a priori meaning. A
meaningful formulation requires a velocity operator, the natural candidate
being $\boldsymbol{v}_{y}=i[\boldsymbol{H},\widehat{y}]$ with $\widehat{y}$
the multiplication operator by $y$, together with the corresponding current
$\boldsymbol{j}_{y}=e\boldsymbol{v}_{y}$. Two questions are then open: whether
$\boldsymbol{v}_{y}$ is well defined on the JR states (the operator
$\widehat{y}$ is unbounded and $|y|_{p}$ takes values in $p^{\mathbb{Z}}$),
and whether the two branches $\pi^{-1}(s)=\pm1$ carry opposite expectation
values of $\boldsymbol{j}_{y}$. We expect that the correct $p$-adic substitute
for ``propagation along the edge'' is transport along the scale direction of
the Bruhat-Tits tree whose boundary is $\mathbb{Q}_{p}$, rather than
translation in $y$.

\section{\label{Section_physics}The physical content of the model}

The analysis above has a physical reading, which we state separately because
it is the reason for studying the model at all. The question behind $p$-adic
quantum mechanics is whether the hypothesis that physical space is totally
disconnected at short distances has consequences that a local theory cannot
produce. The $p$-adic Jackiw-Rebbi model exhibits three such consequences:
the stationary states are strictly localized, the momentum is quantized
geometrically, and the masses that admit an interface state at $E=0$ obey a
selection rule. We present them in the next three subsections, we then add an
operational statement, namely that the model is realizable at every finite
level of resolution, and we close by recording what the model does not give.

\subsection{Every stationary state is strictly localized}

We first record a fact that is implicit in Section 2.8. The functions
$\Phi_{rnj}$, with $r\in\mathbb{Z}$, $n\in\mathbb{Q}_{p}/\mathbb{Z}_{p}$ and
$j\in\mathbb{F}_{p}^{\times}$, form a complete orthonormal system of
$L^{2}(\mathbb{Q}_{p})$: they are the wavelets introduced by Kozyrev,
\cite{Kozyrev}, see also \cite{Zuniga-Textbook}, and by construction each of
them vanishes outside a single ball. By (\ref{Eigenfunctions}) they are
eigenfunctions of $\widetilde{\boldsymbol{D}}_{x}$. The free $p$-adic Dirac
Hamiltonian is therefore diagonalized by a basis of compactly supported
functions, which has the following consequences.

\begin{theorem}
\label{Thm_localization}Let $m>0$ be a constant, and let $\boldsymbol{H}%
_{\mathrm{sa}}=v\widetilde{\boldsymbol{D}}_{x}\sigma_{x}+mv^{2}\sigma_{z}$ be
the self-adjoint $p$-adic Dirac Hamiltonian of (\ref{Two_Hamiltonians}) with
that constant mass. For
$r\in\mathbb{Z}$ and $j\in\mathbb{F}_{p}^{\times}$ put $\lambda_{rj}=\pi
^{-1}\left(  j\right)  p^{1-r}$, and let $w_{rj}^{\pm}\in\mathbb{C}^{2}$ be
the two eigenvectors of the matrix $v\lambda_{rj}\sigma_{x}+mv^{2}\sigma_{z}$. Then:

\begin{enumerate}
\item[(i)] the functions $\Phi_{rnj}\otimes w_{rj}^{\pm}$ form a complete
orthonormal system of $L^{2}(\mathbb{Q}_{p})\otimes\mathbb{C}^{2}$, and each
of them vanishes outside the ball $p^{-r}n+p^{-r}\mathbb{Z}_{p}$;

\item[(ii)] $\boldsymbol{H}_{\mathrm{sa}}$ has pure point spectrum, and its eigenvalues
are%
\begin{equation}
E_{k}^{\pm}=\pm\sqrt{m^{2}v^{4}+v^{2}p^{2k}}\text{, \ \ }k\in\mathbb{Z}%
\text{,} \label{energy_ladder}%
\end{equation}
each of infinite multiplicity;

\item[(iii)] $\sigma\left(  \boldsymbol{H}_{\mathrm{sa}}\right)  =\left\{  E_{k}^{\pm
};k\in\mathbb{Z}\right\}  \cup\left\{  \pm mv^{2}\right\}  $, and the two
points $\pm mv^{2}$ belong to the essential spectrum but are not eigenvalues.
\end{enumerate}
\end{theorem}

\begin{proof}
(i) Orthonormality and completeness are Kozyrev's theorem, \cite{Kozyrev}; the
support statement follows from the definition of $\Phi_{rnj}$, which vanishes
unless $\left\vert p^{r}x-n\right\vert _{p}\leq1$. By (\ref{Eigenfunctions}),
$\widetilde{\boldsymbol{D}}_{x}\Phi_{rnj}=\lambda_{rj}\Phi_{rnj}$, so
$\boldsymbol{H}_{\mathrm{sa}}$ leaves the two-dimensional space $\mathbb{C}^{2}%
\otimes\Phi_{rnj}$ invariant and acts on it as the matrix $v\lambda_{rj}%
\sigma_{x}+mv^{2}\sigma_{z}$, whose eigenvalues are $\pm\sqrt{m^{2}v^{4}%
+v^{2}\lambda_{rj}^{2}}$ by the computation in the proof of Proposition
\ref{Prop_gap}. (ii) Since $\left\vert \lambda_{rj}\right\vert =p^{1-r}$
depends only on $r$, and $r$ runs over $\mathbb{Z}$, the eigenvalues are those
of (\ref{energy_ladder}); each has infinite multiplicity because $n$ runs over
the countably infinite set $\mathbb{Q}_{p}/\mathbb{Z}_{p}$ and $j$ over
$\mathbb{F}_{p}^{\times}$. (iii) The spectrum of a self-adjoint operator is
closed, and the set in (ii) accumulates exactly at $\pm mv^{2}$, which
correspond to $r\rightarrow+\infty$. A non-zero vector in the kernel of
$\boldsymbol{H}_{\mathrm{sa}}\mp mv^{2}$ would satisfy $\widetilde{\boldsymbol{D}}%
_{x}\psi=0$, which is impossible because the $\Phi_{rnj}$ are complete and
their eigenvalues are non-zero.
\end{proof}

\begin{remark}
\label{Remark_Archimedean_contrast}None of this is available in the
Archimedean case. The free Dirac operator on $L^{2}(\mathbb{R}^{3}%
)\otimes\mathbb{C}^{4}$ has purely absolutely continuous spectrum $\left(
-\infty,-mc^{2}\right]  \cup\left[  mc^{2},\infty\right)  $ and no eigenvalues
at all, so it has no stationary states in the Hilbert space, \cite[Section
1.4]{Thaller}; and no non-trivial solution of the free Dirac equation can have
compact support at two different times, \cite[Section 1.8]{Thaller}, which is
the analytic expression of Einstein causality. In the $p$-adic model the free
Hamiltonian has, by contrast, pure point spectrum, and its eigenfunctions form
a complete orthonormal system, so that every state of the Hilbert space is a
superposition of stationary states; and each of these stationary states is
strictly localized, in the strong sense of vanishing outside a ball. This is
the sharpest physical difference between the two frameworks, and it is a
consequence of the ultrametric geometry rather than of the choice of
Hamiltonian: it holds for the free particle, before any mass profile is
introduced, and it holds for $\boldsymbol{H}_{\mathrm{nh}}$ as well, since
that operator is block-diagonalized by the same basis.
\end{remark}

This is also the point at which the $p$-adic framework touches the discussion
of localization in relativistic quantum mechanics. In standard quantum
mechanics, on a configuration space that locally looks like $\mathbb{R}%
\times\mathbb{R}^{N}$, the existence of localized wavefunctions is
incompatible with causality: the Newton-Wigner scheme, \cite{Newton-Wigner},
predicts a superluminal spreading, \cite{Felimng et al}, by the Hegerfeldt
theorem, \cite{Hegerfeldt,Hegerfeldt et al}, and for this reason it is
not regarded as fundamental, while relativistic quantum field theory forbids
faster-than-light communication, which is the content of the no-communication
theorem of Eberhard and Ross, \cite{Eberhard et al}. Whether the corresponding
statements hold when $\mathbb{R}\times\mathbb{R}^{3}$ is replaced by
$\mathbb{R}\times\mathbb{Q}_{p}^{3}$ is an open problem, see
\cite{Zuniga-QM-2}, \cite{Zuniga-Galindo-ultimo}. As explained in the
introduction, the strict localization established here contradicts none of
these results, since the Archimedean argument rests on the connectedness of
the line; but it makes the question concrete.

\subsection{\label{Section_momentum}Momentum is quantized, and the two
theories part at high energy}

By Lemma \ref{Lemma_selfadjoint} the momentum operator of the model is
$\widehat{\boldsymbol{p}}=\widetilde{\boldsymbol{D}}_{x}$, and by
(\ref{Eigenfunctions}) its spectrum is%
\[
\sigma\left(  \widehat{\boldsymbol{p}}\right)  =\left\{  \pm p^{k}%
;k\in\mathbb{Z}\right\}  \cup\left\{  0\right\}  \text{.}%
\]
Momentum therefore takes its values on a geometric ladder rather than on a
continuum, and the lengths associated with it, namely the inverses of the
momenta, are the radii of balls, that is, powers of $p$. This is the precise
sense in which the prime carries inverse length units, as discussed in Section
\ref{Section_Compare}: the discreteness of space is not an approximation
imposed on the model but a property of its spectrum.

The energies (\ref{energy_ladder}) make the physical consequence quantitative.
They are indexed by $k\in\mathbb{Z}$, which is related to the scale $r$ of the
wavelet by $k=1-r$, see (\ref{Eigenfunctions}); the momentum of the level
$E_{k}^{\pm}$ is $p^{k}$ in absolute value, so that $k\rightarrow-\infty$, that
is $r\rightarrow+\infty$, is the low-momentum regime, and $k\rightarrow
+\infty$, that is $r\rightarrow-\infty$, the high-momentum one.

For $\left\vert E\right\vert $ close to $mv^{2}$, that is for $k\rightarrow
-\infty$, the expansion of the square root in (\ref{energy_ladder}) in powers
of the small quantity $p^{2k}/\left(  m^{2}v^{2}\right)  $ gives%
\[
E_{k}^{+}-mv^{2}=mv^{2}\left(  \sqrt{1+\frac{p^{2k}}{m^{2}v^{2}}}-1\right)
\simeq\frac{p^{2k}}{2m}\text{,}%
\]
so the levels accumulate at the gap edge geometrically fast and the spectrum
is indistinguishable from the continuous spectrum of the Archimedean model at
any finite resolution. For $\left\vert E\right\vert $ well above $mv^{2}$, on
the other hand, the term $v^{2}p^{2k}$ dominates in (\ref{energy_ladder}), so
that $E_{k}^{+}\simeq vp^{k}$ and%
\[
\frac{E_{k+1}^{+}}{E_{k}^{+}}\longrightarrow p\text{, \ \ }k\rightarrow
+\infty\text{,}%
\]
so consecutive levels differ by a fixed factor. The two regimes are visible in
Figure \ref{Fig_dispersion}(a). The $p$-adic structure of space is thus
invisible at low energy and dominant at high energy, which is the behaviour
one expects of a hypothesis about the structure of space at the Planck length,
and it is the reason why a model of this kind is not in conflict with
low-energy experiments.

The momentum ladder is a property of $\widetilde{\boldsymbol{D}}_{x}$, and is
therefore the same for the two models of (\ref{Two_Hamiltonians}); the
energies are not. On a ball where the mass is constant, $\boldsymbol{H}%
_{\mathrm{nh}}$ acts on $\mathbb{C}^{2}\otimes\Theta$ as the matrix
$\boldsymbol{h}\left(  \lambda\right)  $ of Section \ref{Section_EP}, whose
eigenvalues are $E=\pm\sqrt{m^{2}v^{4}-v^{2}\lambda^{2}}$, with the sign under
the square root reversed. For $k\rightarrow-\infty$ this gives%
\[
mv^{2}-E_{k}^{+}\simeq\frac{p^{2k}}{2m}\text{,}%
\]
so that the levels accumulate at the gap edge at the same geometric rate,
although from inside the gap. For $p^{k}>mv$, on the other hand, the
expression under the square root is negative and the energies are purely
imaginary, $E_{k}^{\pm}\simeq\pm ivp^{k}$; then $E_{k+1}^{\pm}/E_{k}^{\pm}\rightarrow p$
as $k\rightarrow+\infty$, that is, consecutive levels again differ by the
factor $p$. The high-energy separation of the two
theories is thus the same in both models, while the threshold $p^{k}=mv$,
where the two eigenvalues coalesce, is the exceptional point of Section
\ref{Section_EP} and has no Archimedean counterpart.

\subsection{The mass at an interface is quantized by the geometry}

The third effect is the one described in Section \ref{Section_EP}. In the
Archimedean Jackiw-Rebbi model a zero-energy bound state appears at the wall
for \emph{every} pair of positive masses, and its decay length $1/(mv)$ may
take any positive value. In the $p$-adic model the decay lengths available to
a bound state are the radii of balls, that is, powers of $p$, and consequently
a zero-energy interface state exists if and only if the masses satisfy the
quantization condition (\ref{mass_quantization}). The geometry of the space
selects which masses can support such a state; no Archimedean mechanism does
this. We stress that this effect belongs to the non-Hermitian model, so that,
physically, it describes an interface in an open system rather than in a
conservative one; for the self-adjoint model the corresponding question is the
open problem of Remark \ref{Remark_origin}.

\subsection{\label{Section_CTQW}The model is realizable at every finite level}

The three effects above are properties of an operator acting on $L^{2}%
(\mathbb{Q}_{p})$, and a natural objection is that such an operator is an
abstraction: the configuration space $\mathbb{Q}_{p}$ is not accessible to
measurement, so it is unclear in what sense its dynamics is physical. We
answer this by showing that the dynamics is the limit of quantum walks on
finite graphs, which connects the present model with the continuous-time
quantum walks studied in \cite{Zuniga-CTQW}.

For integers $L,l\geq1$, put $N=L+l$, and set%
\[
G_{-L,l}=p^{-L}\mathbb{Z}_{p}/p^{l}\mathbb{Z}_{p}\text{,}%
\]
which is a quotient group. We use%
\[
a=a_{-L}p^{-L}+a_{-L+1}p^{-L+1}+\ldots+a_{0}+a_{1}p+\ldots+a_{l-1}p^{l-1}%
\]
as a set of representatives for the elements of $G_{-L,l}$. Given $a\in
G_{-L,l}$, the set $B_{-l}(a)=a+p^{l}\mathbb{Z}_{p}$ is the ball of radius
$p^{-l}$ centred at $a$, and we denote by $\Omega\left(  p^{l}\left\vert
x-a\right\vert _{p}\right)  $ its characteristic function. Then $\left\{
\Omega\left(  p^{l}\left\vert x-a\right\vert _{p}\right)  \right\}  _{a\in
G_{-L,l}}$ is a set of linearly independent functions with $p^{N}%
=\#G_{-L,l}$ elements, and we denote by $\mathcal{D}_{l}^{-L}(\mathbb{Q}_{p})$
the space spanned by them. The elements of $\mathcal{D}_{l}^{-L}%
(\mathbb{Q}_{p})$ are the functions on $\mathbb{Q}_{p}$ that are locally
constant of exponent $l$ and supported in $B_{L}$. The elements of $G_{-L,l}$
are the leaves of a finite rooted tree of depth $N$, the ultrametric distance
between two leaves being determined by the level at which their branches
separate.

\begin{theorem}
\label{Prop_CTQW}Let $W_{L,l}\subset\mathcal{D}_{l}^{-L}(\mathbb{Q}_{p})$ be the linear span of the
wavelets $\Phi_{rnj}$ whose support is contained in $B_{L}$, with $1-l\leq
r\leq L$. Let $m>0$ be constant and let $\boldsymbol{H}_{\mathrm{sa}}$ be as in Theorem
\ref{Thm_localization}. Then:

\begin{enumerate}
\item[(i)] $W_{L,l}=\left\{  f\in\mathcal{D}_{l}^{-L}(\mathbb{Q}_{p});\int_{\mathbb{Q}_{p}}f\left(
x\right)  dx=0\right\}  $, and $\dim W_{L,l}=p^{N}-1$;

\item[(ii)] $W_{L,l}$ is invariant under $\widetilde{\boldsymbol{D}}_{x}$, and
$\boldsymbol{H}_{\mathrm{sa}}$ leaves $W_{L,l}\otimes\mathbb{C}^{2}$ invariant; the
restriction $\boldsymbol{H}^{L}:=\boldsymbol{H}_{\mathrm{sa}}|_{W_{L,l}\otimes
\mathbb{C}^{2}}$ is a Hermitian matrix of size $2\left(  p^{N}-1\right)  $,
which is diagonal in the basis of Theorem \ref{Thm_localization};

\item[(iii)] the spaces $W_{L,l}$ increase with $L$ and $l$, and their union
is a core for $\boldsymbol{H}_{\mathrm{sa}}$; consequently $\boldsymbol{H}%
^{L}\rightarrow\boldsymbol{H}_{\mathrm{sa}}$ in the strong resolvent sense, and
$e^{-it\boldsymbol{H}^{L}}\rightarrow e^{-it\boldsymbol{H}_{\mathrm{sa}}}$ strongly,
uniformly for $t$ in compact intervals.
\end{enumerate}
\end{theorem}

\begin{proof}
(i) Each $\Phi_{rnj}$ is the product of a non-trivial additive character with
the characteristic function of a ball, so its integral vanishes, and the
restriction $1-l\leq r$ is exactly the condition that $\Phi_{rnj}$ be locally
constant of exponent $l$, and the requirement $\mathrm{supp}\,\Phi
_{rnj}\subset B_{L}$, which forces $r\leq L$, places its support inside
$B_{L}$. Counting these wavelets gives $\sum_{r=1-l}^{L}\left(  p-1\right)
p^{L-r}=p^{N}-1$, which is the dimension of the space of the elements of
$\mathcal{D}_{l}^{-L}(\mathbb{Q}_{p})$ with zero mean; since $W_{L,l}$ is contained in that space and has the same
dimension, the two coincide. (ii) is immediate from (\ref{Eigenfunctions}),
the $\Phi_{rnj}$ being eigenfunctions of $\widetilde{\boldsymbol{D}}_{x}$, and
from the fact that the mass term acts on the second factor only. (iii) The
union of the $W_{L,l}$ is the algebraic span of the complete orthonormal
system $\left\{  \Phi_{rnj}\right\}  $ of Theorem \ref{Thm_localization},
which is a core for any self-adjoint operator diagonalized by that system, and
the mass term is a bounded perturbation. The convergence statements then
follow from the standard criteria for strong resolvent convergence and for the
associated unitary groups, see \cite[Section VIII.7]{Reed-Simon}.
\end{proof}

Theorem \ref{Prop_CTQW} admits the following reading. The matrix
$\boldsymbol{H}^{L}$ is indexed by the $p^{N}$ elements of $G_{-L,l}$, the
leaves of a finite rooted tree of depth $N$, up to the one-dimensional
direction discussed below, and its off-diagonal entries
are the hopping amplitudes supplied by the kernel of $\widetilde
{\boldsymbol{D}}_{x}$, which are long-ranged and decay with the ultrametric
distance. The one-parameter group $e^{-it\boldsymbol{H}^{L}}$ is therefore a
continuous-time quantum walk on that vertex set, in the sense of
\cite{Zuniga-CTQW}, with an internal two-component degree of freedom; and by
(iii) the $p$-adic Dirac dynamics is the limit of these walks. In this sense
the model, although formulated on a space that is not directly accessible to
measurement, prescribes a dynamics that is in principle implementable at every
finite level of resolution, on a system with $p^{N}$ sites.

\begin{remark}
We now discuss two relevant matters. First, the walks constructed in
\cite{Zuniga-CTQW} are generated by \emph{bounded} non-local operators with
integrable kernels, a hypothesis under which the walk is defined on the whole
space. Boundedness is not needed in order to obtain a continuous-time quantum
walk: the construction rests on general principles, and unbounded operators of
Schr\"{o}dinger type are treated in this way in \cite{Zuniga-QM-2} and
\cite{Zuniga-Chacon}. Here $\widetilde{\boldsymbol{D}}_{x}$ is unbounded, so
that the correspondence obtained above holds at each finite level and the
limit is strong rather than uniform. Second, Theorem \ref{Prop_CTQW} is stated for a
constant mass. For the Jackiw-Rebbi profile (\ref{function_m(x)})
multiplication by $m(\cdot)$ does not preserve $W_{L,l}$, because that profile
is not locally constant at the origin (Remark \ref{Remark_origin}); the
finite-level operators are then compressions rather than restrictions, and the
corresponding convergence requires a Trotter-Kato argument that we do not
develop here.
\end{remark}

\subsection{\label{Section_nh_physics}The non-Hermitian model}

The statements above were made for $\boldsymbol{H}_{\mathrm{sa}}$ with a
constant mass, except for the quantization of the mass, which is a statement
about $\boldsymbol{H}_{\mathrm{nh}}$ from the start. We record here what
becomes of them for the non-Hermitian operator.

Strict localization survives in the same form. The wavelets $\Phi_{rnj}$ are
eigenfunctions of $\widetilde{\boldsymbol{D}}_{x}$, so $\boldsymbol{H}%
_{\mathrm{nh}}$ leaves each two-dimensional space $\mathbb{C}^{2}\otimes
\Phi_{rnj}$ invariant and acts on it as the matrix $\boldsymbol{h}\left(
\lambda_{rj}\right)  $ of Section \ref{Section_EP}, whose eigenvalues are
$\pm\sqrt{m^{2}v^{4}-v^{2}\lambda_{rj}^{2}}$. These eigenvalues are distinct,
and the matrix is diagonalizable, unless $v\left\vert \lambda_{rj}\right\vert
=mv^{2}$, which occurs only when the mass satisfies the quantization condition
(\ref{mass_quantization}); away from those scales the eigenvectors of
$\boldsymbol{h}\left(  \lambda_{rj}\right)  $ produce a complete family of
eigenfunctions of $\boldsymbol{H}_{\mathrm{nh}}$, each of them vanishing
outside a ball, with the orthogonality of Theorem \ref{Thm_localization}
replaced by the biorthogonality of the eigenvectors of a non-normal matrix.
The momentum spectrum is the same geometric ladder, since it is a property of
$\widetilde{\boldsymbol{D}}_{x}$ alone, and the energies attached to it were
described in Section \ref{Section_momentum}.

The finite-level description also survives, with one essential difference.
The space $W_{L,l}$ of Theorem \ref{Prop_CTQW} is invariant under
$\widetilde{\boldsymbol{D}}_{x}$, so $\boldsymbol{H}_{\mathrm{nh}}$ leaves
$W_{L,l}\otimes\mathbb{C}^{2}$ invariant as well, and the restriction
$\boldsymbol{H}_{\mathrm{nh}}^{L}$ is a matrix of size $2\left(
p^{N}-1\right)  $, indexed by the same vertex set $G_{-L,l}$, which is no
longer Hermitian. The one-parameter group $e^{-it\boldsymbol{H}_{\mathrm{nh}%
}^{L}}$ is therefore a continuous-time walk that is not unitary: at each
finite level the model describes an open system, whose amplitudes grow or
decay at the rates given by the imaginary parts of the eigenvalues. This is
not an accident of the construction. By Stone's theorem, a strongly continuous
one-parameter group of unitary operators on a Hilbert space has a self-adjoint
generator, and conversely, see \cite[Theorem VIII.8]{Reed-Simon}; a
Hamiltonian that is not self-adjoint cannot therefore generate a unitary
dynamics, at any level of resolution. The limit
statement of Theorem \ref{Prop_CTQW}(iii) has no counterpart here, and for a
stronger reason: by the formula above the imaginary parts of the eigenvalues
are unbounded as $r\rightarrow-\infty$, so that $\boldsymbol{H}_{\mathrm{nh}}$
generates no one-parameter group of bounded operators on $L^{2}(\mathbb{Q}%
_{p})\otimes\mathbb{C}^{2}$, unitary or not, and the finite-level walks
converge to no such group. The realizability of the
non-Hermitian model is thus a statement about each finite level, which is
where an experiment would be performed, and not about the limit.

\subsection{What the model does not give}

It is equally important to say what does not survive the passage to a totally
disconnected space. The Archimedean Jackiw-Rebbi zero mode is protected: the
number of chiral zero modes at a wall equals the jump of a winding number
across it, and no deformation preserving the chiral symmetry and the gap can
remove it. Over $\mathbb{Q}_{p}$ that winding number does not exist, by
Proposition \ref{Prop_nogo} below, and neither does the index argument that
pins the mode to zero energy. The fermion-number argument that assigns the
charge $\pm\frac{1}{2}$ to the vacuum rests on the same ingredient and is
likewise unavailable. What survives is the mechanism, namely that a change of
sign of the mass binds states at the interface; what does not survive is the topology.

Taken together, these statements are, in our view, the physical content of the
model: $p$-adic quantum mechanics, although it is a non-local theory,
reproduces the physical picture of the Archimedean model, and it realizes that
picture by mechanisms that have no Archimedean counterpart. The picture is the
familiar one: a gapped spectrum, symmetric about zero energy, with bound
states at an interface where the mass changes sign, and indistinguishable from
the Archimedean spectrum at low energy. The mechanisms are not: strictly
localized stationary states, a geometrically quantized momentum, and a
selection rule for the masses that admit a zero-energy interface state. The
model prescribes, in addition, at each finite level of resolution, a dynamics
that is in principle implementable; and it loses the topological protection
that the Archimedean model derives from the connectedness of the line.

\section{\label{Section_invariants}Symmetry classes and the absence of
$p$-adic band invariants}

\subsection{Symmetry class}

By Lemma \ref{Lemma_Gauss}, the operator $\widetilde{\boldsymbol{D}}_{x}$ is
odd under complex conjugation when $p\equiv3$ $(\operatorname{mod}4)$ and even
when $p\equiv1$ $(\operatorname{mod}4)$. By interpreting $k\left(  y\right)  =\pi\left(
y\right)  \Gamma_{p}\left(  -1,\pi\right)  ^{-1}\left\vert y\right\vert
_{p}^{-2}$ as the kernel of $\widetilde{\boldsymbol{D}}_{x}$, see
Section \ref{Section_Vladimirov}, the dichotomy just mentioned can also be
read off from that kernel, which is how we record it in the table below. Since $\pi$ takes the values $\pm1$ and $\left\vert
y\right\vert _{p}^{-2}$ is positive, complex conjugation replaces $k$
by%
\[
\overline{k}=\frac{\Gamma_{p}\left(  -1,\pi\right)  }{\overline{\Gamma
_{p}\left(  -1,\pi\right)  }}\text{ }k\text{,}%
\]
so that the second identity of Lemma \ref{Lemma_Gauss}(i) says precisely that
$\Gamma_{p}\left(  -1,\pi\right)  $, and with it $k$, is real when
$\pi\left(  -1\right)  =1$ and purely imaginary when $\pi\left(  -1\right)
=-1$. This is what happens, by the value (\ref{Gauss_sum}) of the Gauss sum:
$\Gamma_{p}\left(  -1,\pi\right)  $ is a positive multiple of
$\sqrt{p}$ when $p\equiv1$ $(\operatorname{mod}4)$ and of $i\sqrt{p}$ when
$p\equiv3$ $(\operatorname{mod}4)$. For the self-adjoint model the antiunitary symmetries are governed by the same
dichotomy, as follows.

\begin{lemma}
\label{Lemma_AZ}Let $m(\cdot)$ be a real, bounded, locally constant function
which is not identically zero, and let $\boldsymbol{H}_{\mathrm{sa}%
}=v\widetilde{\boldsymbol{D}}_{x}\sigma_{x}+m(x)v^{2}\sigma_{z}$. We call an
antiunitary operator $\mathcal{T}$ a \emph{time-reversal symmetry} if
$\mathcal{T}\boldsymbol{H}_{\mathrm{sa}}\mathcal{T}^{-1}=\boldsymbol{H}%
_{\mathrm{sa}}$, an antiunitary operator $\mathcal{C}$ a \emph{particle-hole
symmetry} if $\mathcal{C}\boldsymbol{H}_{\mathrm{sa}}\mathcal{C}^{-1}%
=-\boldsymbol{H}_{\mathrm{sa}}$, and a unitary operator $\Gamma$ a
\emph{chiral symmetry} if $\Gamma\boldsymbol{H}_{\mathrm{sa}}\Gamma
^{-1}=-\boldsymbol{H}_{\mathrm{sa}}$. Then the following assertions hold.

(i) $\Gamma=\sigma_{y}$ is a chiral symmetry, for every prime $p>2$.

(ii) If $p\equiv3$ $(\operatorname{mod}4)$, the only antiunitary symmetries of
the form $\boldsymbol{UK}$, with $\boldsymbol{U}$ a unitary $2\times2$ matrix,
are, up to a phase, $\mathcal{T}=\sigma_{z}\boldsymbol{K}$ and $\mathcal{C}%
=\sigma_{x}\boldsymbol{K}$, and $\mathcal{T}^{2}=\mathcal{C}^{2}%
=\boldsymbol{1}$.

(iii) If $p\equiv1$ $(\operatorname{mod}4)$, they are, again up to a phase,
$\mathcal{T}=\boldsymbol{K}$ and $\mathcal{C}=\sigma_{y}\boldsymbol{K}$, and
$\mathcal{T}^{2}=\boldsymbol{1}$, $\mathcal{C}^{2}=-\boldsymbol{1}$.

(iv) In both cases $\mathcal{TC}$ is a scalar multiple of $\Gamma$.
\end{lemma}

\begin{proof}
Set $\varepsilon=\pi\left(  -1\right)  $, so that $\varepsilon=-1$ for
$p\equiv3$ $(\operatorname{mod}4)$ and $\varepsilon=1$ for $p\equiv1$
$(\operatorname{mod}4)$, by (\ref{Euler_criterion}). The matrices $\sigma_{x}%
$, $\sigma_{y}$, $\sigma_{z}$ commute with $\widetilde{\boldsymbol{D}}_{x}$
and with the multiplication by $m\left(  x\right)  $, since these act on the
scalar factor. Assertion (i) is then immediate from $\sigma_{y}\sigma
_{x}\sigma_{y}=-\sigma_{x}$ and $\sigma_{y}\sigma_{z}\sigma_{y}=-\sigma_{z}$.

Let now $\boldsymbol{U}$ be a unitary $2\times2$ matrix. The matrices
$\sigma_{x}$, $\sigma_{z}$ have real entries, so that $\boldsymbol{K}%
\sigma_{x}\boldsymbol{K}=\sigma_{x}$ and $\boldsymbol{K}\sigma_{z}%
\boldsymbol{K}=\sigma_{z}$, while $\boldsymbol{K}\sigma_{y}\boldsymbol{K}%
=-\sigma_{y}$. The mass $m\left(  x\right)  $ is real, so $\boldsymbol{K}$
commutes with the multiplication by it, and $\boldsymbol{K}\widetilde
{\boldsymbol{D}}_{x}\boldsymbol{K}=\varepsilon\widetilde{\boldsymbol{D}}_{x}$
by Lemma \ref{Lemma_Gauss}(i). Consequently%
\[
\left(  \boldsymbol{UK}\right)  \boldsymbol{H}_{\mathrm{sa}}\left(
\boldsymbol{UK}\right)  ^{-1}=\varepsilon v\widetilde{\boldsymbol{D}}%
_{x}\left(  \boldsymbol{U}\sigma_{x}\boldsymbol{U}^{-1}\right)  +m\left(
x\right)  v^{2}\left(  \boldsymbol{U}\sigma_{z}\boldsymbol{U}^{-1}\right)
\text{.}%
\]
The operator $\widetilde{\boldsymbol{D}}_{x}$ is not a multiplication
operator, hence it is linearly independent of the multiplication by $m\left(
x\right)  $, and the two terms can be compared separately: $\boldsymbol{UK}$
is a time-reversal symmetry if and only if%
\begin{equation}
\varepsilon\boldsymbol{U}\sigma_{x}\boldsymbol{U}^{-1}=\sigma_{x}\text{,
\ \ }\boldsymbol{U}\sigma_{z}\boldsymbol{U}^{-1}=\sigma_{z}\text{,}
\label{AZ_conditions}%
\end{equation}
and a particle-hole symmetry if and only if the right-hand sides of
(\ref{AZ_conditions}) are replaced by $-\sigma_{x}$ and $-\sigma_{z}$.

In the first case $\boldsymbol{U}$ commutes with $\sigma_{z}$, which has two
distinct eigenvalues, so $\boldsymbol{U}=\mathrm{diag}\left(  u_{1}%
,u_{2}\right)  $ with $\left\vert u_{1}\right\vert =\left\vert u_{2}%
\right\vert =1$, and%
\[
\boldsymbol{U}\sigma_{x}\boldsymbol{U}^{-1}=\left[
\begin{array}
[c]{cc}%
0 & u_{1}u_{2}^{-1}\\
u_{2}u_{1}^{-1} & 0
\end{array}
\right]  \text{,}%
\]
so that the first condition in (\ref{AZ_conditions}) reads $u_{1}u_{2}%
^{-1}=\varepsilon$. Thus $\boldsymbol{U}$ is a phase times $\boldsymbol{1}$
when $\varepsilon=1$, and a phase times $\sigma_{z}$ when $\varepsilon=-1$. In
the second case $\boldsymbol{U}$ anticommutes with $\sigma_{z}$, hence it is
off-diagonal, $\boldsymbol{U}=\left[
\begin{array}
[c]{cc}%
0 & b\\
c & 0
\end{array}
\right]  $ with $\left\vert b\right\vert =\left\vert c\right\vert =1$, and the
same computation gives $\boldsymbol{U}\sigma_{x}\boldsymbol{U}^{-1}$ with
off-diagonal entries $bc^{-1}$ and $cb^{-1}$; the condition $\varepsilon
\boldsymbol{U}\sigma_{x}\boldsymbol{U}^{-1}=-\sigma_{x}$ reads $bc^{-1}%
=-\varepsilon$. Thus $\boldsymbol{U}$ is a phase times $\sigma_{x}$ when
$\varepsilon=-1$, and a phase times $\sigma_{y}$ when $\varepsilon=1$, because
$\left[
\begin{array}
[c]{cc}%
0 & -1\\
1 & 0
\end{array}
\right]  =-i\sigma_{y}$. This proves the two lists in (ii) and (iii).

For the squares, $\left(  e^{i\theta}\boldsymbol{UK}\right)  ^{2}%
=\boldsymbol{U}\overline{\boldsymbol{U}}$, the phase cancelling against its
conjugate, and $\boldsymbol{U}\overline{\boldsymbol{U}}=\boldsymbol{1}$ for
$\boldsymbol{U}\in\left\{  \boldsymbol{1},\sigma_{x},\sigma_{z}\right\}  $,
while $\sigma_{y}\overline{\sigma_{y}}=-\sigma_{y}^{2}=-\boldsymbol{1}$.
Finally $\sigma_{z}\boldsymbol{K}\sigma_{x}\boldsymbol{K}=\sigma_{z}\sigma
_{x}=i\sigma_{y}$ and $\boldsymbol{K}\sigma_{y}\boldsymbol{K}=-\sigma_{y}$,
which is (iv).
\end{proof}

The operators of Lemma \ref{Lemma_AZ} are not the ones of Lemma
\ref{Lemma_PT}: the latter concerns $\boldsymbol{H}_{\mathrm{nh}}$, whose
kinetic term carries an extra factor $-i$, and this exchanges the two
congruence classes. Lemma \ref{Lemma_AZ} gives the following table.

\begin{center}%
\begin{tabular}
[c]{|l|l|l|}\hline
Property of the model & $p\equiv3$ $(\operatorname{mod}4)$ & $p\equiv1$
$(\operatorname{mod}4)$\\\hline
Kernel of $\widetilde{\boldsymbol{D}}_{x}$ & purely imaginary & real\\\hline
Analogue & $-i\partial_{x}$, odd under $\boldsymbol{K}$ & real symmetric
hopping, even under $\boldsymbol{K}$\\\hline
Chiral symmetry & $\Gamma=\sigma_{y}$ & $\Gamma=\sigma_{y}$\\\hline
Antiunitary symmetries & $\mathcal{T}=\sigma_{z}\boldsymbol{K}$, $\mathcal{C}%
=\sigma_{x}\boldsymbol{K}$ & $\mathcal{T}=\boldsymbol{K}$, $\mathcal{C}%
=\sigma_{y}\boldsymbol{K}$\\\hline
Class & BDI $\left(  \mathcal{T}^{2}=\mathcal{C}^{2}=\boldsymbol{1}\right)  $
& CI $\left(  \mathcal{T}^{2}=\boldsymbol{1}\text{, }\mathcal{C}%
^{2}=-\boldsymbol{1}\right)  $\\\hline
\end{tabular}

\end{center}

The labels in the last row are those of the Altland-Zirnbauer
classification, \cite{Kitaev-table,Chiu et al}, in which a chiral
symmetry with no antiunitary symmetry gives the class AIII; a chiral symmetry
together with $\mathcal{T}^{2}=\mathcal{C}^{2}=\boldsymbol{1}$ gives BDI; and
a chiral symmetry together with $\mathcal{T}^{2}=\boldsymbol{1}$,
$\mathcal{C}^{2}=-\boldsymbol{1}$ gives CI. Thus a quadratic-residue condition
on the prime selects the symmetry class of the model, which is BDI for
$p\equiv3$ $(\operatorname{mod}4)$ and CI for $p\equiv1$ $(\operatorname{mod}%
4)$. We are not aware of any Archimedean analogue of this phenomenon,
and we regard it as one of the genuinely new features of the non-Archimedean
setting; a systematic analysis, including multiplicative characters of order
larger than two, is left for future work.

\subsection{\label{Section_BZ}The Brillouin zone in the $p$-adic framework}

To go further, we need to provide a definition of Chern numbers over
$\mathbb{Q}_{p}$. This requires saying first what plays the role of the
Brillouin zone, and what the invariants of topological band theory are. We
recall both, since the two communities involved use different languages.

\subsubsection{What plays the role of the Brillouin zone}

In the theory of crystals the configuration space is a lattice, say
$\mathbb{Z}^{d}\subset\mathbb{R}^{d}$, and a Hamiltonian commuting with the
translations by $\mathbb{Z}^{d}$ is diagonalized by the Fourier transform of
that group. The Fourier transform of a locally compact abelian group $G$ is
defined on the \textit{Pontryagin dual}%
\[
\widehat{G}=\left\{  \text{continuous homomorphisms }G\rightarrow S\right\}
\text{, \ \ }S=\left\{  z\in\mathbb{C};\left\vert z\right\vert =1\right\}
\text{,}%
\]
see, e.g., \cite{Rudin-groups,Hewitt-Ross}. For $G=\mathbb{Z}^{d}$ one
has $\widehat{G}=\mathbb{T}^{d}$, the $d$-dimensional torus, and this torus is
what physicists call the \textit{Brillouin zone} (BZ). A Hamiltonian
commuting with the translations is then a continuous family of finite matrices
$H\left(  k\right)  $ indexed by $k\in\mathrm{BZ}$, and all the invariants of
topological band theory are invariants of that family, hence invariants of
maps defined on $\mathrm{BZ}$,
\cite{Kitaev-table,Chiu et al}, \cite{Prodan-Schulz-Baldes}.

The same definition applies verbatim in the $p$-adic framework, and this is
the meaning we give to the symbol $\mathrm{BZ}$ in this paper. The group of
translations is now $\mathbb{Q}_{p}$ (or $\mathbb{Q}_{p}^{N}$), and by the
self-duality of $\mathbb{Q}_{p}$ under the pairing $\left(  \xi,x\right)
\mapsto\chi_{p}\left(  \xi x\right)  $, see \cite[Chapter 2]{V-V-Z},
\cite{Taibleson}, we have

\begin{center}%
\begin{tabular}
[c]{|l|l|l|}\hline
Translations by & Brillouin zone $\mathrm{BZ}=\widehat{G}$ &
Description\\\hline
$\mathbb{Q}_{p}^{N}$ & $\mathbb{Q}_{p}^{N}$ & self-dual, locally
compact\\\hline
$\mathbb{Z}_{p}^{N}$ (unit ball) & $\left(  \mathbb{Q}_{p}/\mathbb{Z}%
_{p}\right)  ^{N}$ & discrete, countable\\\hline
$G_{l}=\mathbb{Z}_{p}/p^{l}\mathbb{Z}_{p}$ & $p^{-l}\mathbb{Z}_{p}%
/\mathbb{Z}_{p}$ & finite cyclic, $p^{l}$ elements\\\hline
\end{tabular}

\end{center}

The third line is the Brillouin zone of the finite quotients $G_{l}$. In the
first two cases $\mathrm{BZ}$ is totally disconnected and, in the compact case
$\mathbb{Z}_{p}^{N}$, homeomorphic to a Cantor set; it is never a torus, and
it is never connected.

This is not an abstract remark: the operators of this paper are diagonalized
by exactly this Fourier transform. We take $N=1$, which is the case of the
Hamiltonians (\ref{Two_Hamiltonians}), so that $\mathrm{BZ}=\mathbb{Q}_{p}$;
the two-dimensional model of Section \ref{Section_2D} is the case $N=2$, with
$\mathrm{BZ}=\mathbb{Q}_{p}^{2}$, and the discussion below applies to it
without change. By (\ref{Taiblseon-Vladimirov_operator}), $\widetilde
{\boldsymbol{D}}_{x}$ acts on the character $\chi_{p}\left(  \xi_{0}x\right)
$ by multiplication by $\pi^{-1}\left(  \xi_{0}\right)  \left\vert \xi
_{0}\right\vert _{p}$, indeed%
\[
\widetilde{\boldsymbol{D}}_{x}\chi_{p}\left(  \xi_{0}x\right)  =\mathcal{F}%
_{\xi\rightarrow x}^{-1}(\pi^{-1}\left(  \xi\right)  \left\vert \xi\right\vert
_{p}\delta\left(  \xi-\xi_{0}\right)  )=\pi^{-1}\left(  \xi
_{0}\right)  \left\vert \xi_{0}\right\vert _{p}\chi_{p}\left(
\xi_{0}x\right)  \text{,}%
\]
where $\mathcal{F}_{\xi\rightarrow x}^{-1}$ denotes the Fourier transform on
the space of distributions, see, e.g., \cite{Zuniga-Textbook}. Consequently
the function%
\[
\mathrm{BZ}=\mathbb{Q}_{p}\ni\xi\longmapsto\pi^{-1}\left(  \xi\right)
\left\vert \xi\right\vert _{p}%
\]
is the $p$-adic band function in dimension one: it plays the role that the symbol of
$-i\partial_{x}$, which is the identity function $k\mapsto k$, plays in the
Archimedean case. The two are quite different, since $\pi^{-1}\left(
\xi\right)  \left\vert \xi\right\vert _{p}$ depends on $\xi$ only
through $\left\vert \xi\right\vert _{p}$ and $\pi\left(  \xi\right)  $, and
is therefore far from injective: it is constant on each of the two halves,
given by $\pi\left(  \xi\right)  =\pm1$, of every sphere $\left\vert
\xi\right\vert _{p}=p^{k}$. With a constant mass, the Hamiltonians
(\ref{Two_Hamiltonians}) commute with the translations, and in the Fourier
representation they act as the family of $2\times2$ matrices%
\[
\mathbb{Q}_{p}\ni\xi\longmapsto v\pi^{-1}\left(  \xi\right)  \left\vert
\xi\right\vert _{p}\sigma_{x}+mv^{2}\sigma_{z}\text{,}%
\]
with the factor $-i$ in front of the first term for $\boldsymbol{H}%
_{\mathrm{nh}}$; this is the band structure to which the invariants recalled
below would apply. In dimension two, where $\mathrm{BZ}=\mathbb{Q}_{p}^{2}$,
the operator (\ref{H_2D_sa}) with a constant mass gives in the same way the
family%
\[
\mathbb{Q}_{p}^{2}\ni\left(  \xi_{1},\xi_{2}\right)  \longmapsto v\left(
\pi^{-1}\left(  \xi_{1}\right)  \left\vert \xi_{1}\right\vert _{p}%
\sigma_{x}+\pi^{-1}\left(  \xi_{2}\right)  \left\vert \xi_{2}\right\vert
_{p}\sigma_{y}\right)  +mv^{2}\sigma_{z}\text{,}%
\]
whose two band functions are the coordinate ones. For the mass profile (\ref{function_m(x)}) there
is no band structure at all, since the Hamiltonians do not commute with the
translations; this is the usual situation, the Archimedean Jackiw-Rebbi
Hamiltonian being in the same case.

\subsubsection{\label{Section_invariants_list}The invariants of topological
band theory, and what they require}

Let $X$ be a compact Hausdorff space. We use the following standard objects.

\begin{enumerate}
\item[(a)] $H^{n}\left(  X,\mathbb{Z}\right)  $ denotes \v{C}ech cohomology
with integer coefficients. \v{C}ech theory, rather than singular theory, is
the appropriate one for spaces that are not CW complexes, such as Cantor sets;
see \cite{Eilenberg-Steenrod,Spanier}.

\item[(b)] $K^{0}\left(  X\right)  $ is the Grothendieck group of complex
vector bundles over $X$, and $K^{1}\left(  X\right)  $ is the group of
homotopy classes $\left[  X,U\left(  \infty\right)  \right]  $ of continuous
maps of $X$ into the infinite unitary group; equivalently $K^{j}\left(
X\right)  =K_{j}\left(  C\left(  X\right)  \right)  $, the operator-algebraic
K-theory of the algebra of continuous functions on $X$. See \cite{Atiyah,Blackadar,Rordam et al}.
\end{enumerate}

The invariants used in topological band theory are built from these groups, as
follows. For a gapped family $H\left(  k\right)  $, $k\in\mathrm{BZ}$, the
spectral projection onto the states below the gap defines a complex vector
bundle over $\mathrm{BZ}$, whose class lies in $K^{0}\left(  \mathrm{BZ}%
\right)  $. Which invariant is available depends on the
dimension. In dimension two the first Chern number is the evaluation, on the
fundamental class of $\mathrm{BZ}$, of the first Chern class of that bundle,
an element of $H^{2}\left(  \mathrm{BZ},\mathbb{Z}\right)  $; this is the
integral of the Berry curvature over the Brillouin zone. In dimension one,
for a system with a chiral symmetry, the flattened Hamiltonian is off-diagonal
and its upper-right block is a map $\mathrm{BZ}\rightarrow U\left(  n\right)
$, where $n$ is the rank of that block; its class lies in $K^{1}\left(  \mathrm{BZ}\right)  =\left[
\mathrm{BZ},U\left(  n\right)  \right]  $, and for $\mathrm{BZ}=S$ one has
$K^{1}\left(  S\right)  =\mathbb{Z}$, the integer being the winding number
that counts Jackiw-Rebbi modes. In dimension three the invariant of the chiral
classes is again a winding number, now the degree of such a map on
$\mathbb{T}^{3}$, an element of $H^{3}\left(  \mathrm{BZ},\mathbb{Z}\right)
$. See \cite{Kitaev-table,Chiu et al}, \cite{Prodan-Schulz-Baldes}. In
short: a Chern number needs $H^{2}\left(  \mathrm{BZ},\mathbb{Z}\right)
\neq0$, and a winding number needs $K^{1}\left(  \mathrm{BZ}\right)  \neq0$,
or $H^{3}\left(  \mathrm{BZ},\mathbb{Z}\right)  \neq0$ in dimension three.

\subsubsection{There are no $p$-adic Chern or winding numbers}

\begin{proposition}
\label{Prop_nogo}Let $X$ be a compact, Hausdorff, totally disconnected space;
for instance $\mathbb{Z}_{p}^{N}$, the Brillouin zone of any of the models of
this paper, or any of its finite levels. Then

\begin{enumerate}
\item[(i)] $H^{0}\left(  X,\mathbb{Z}\right)  =C\left(  X,\mathbb{Z}\right)  $
is the group of locally constant integer-valued functions on $X$, and
$H^{n}\left(  X,\mathbb{Z}\right)  =0$ for every $n\geq1$;

\item[(ii)] $K^{0}\left(  X\right)  \cong C\left(  X,\mathbb{Z}\right)  $ and
$K^{1}\left(  X\right)  =0$.
\end{enumerate}

Consequently, over a $p$-adic Brillouin zone every invariant of the kind
recalled in Section \ref{Section_invariants_list} vanishes identically,
whatever the dimension: a Chern number is an element of $H^{2}\left(
X,\mathbb{Z}\right)  $, a winding number in dimension one an element of
$K^{1}\left(  X\right)  $, and a winding number in dimension three an element
of $H^{3}\left(  X,\mathbb{Z}\right)  $, and all these groups are trivial. The
only invariants that can be extracted from the band structure are the rank
functions, that is, counts of states.
\end{proposition}

Both models of this paper are therefore covered, each by the invariant that
its dimension calls for: the winding number of the one-dimensional models
(\ref{Two_Hamiltonians}), and the Chern number of the two-dimensional model of
Section \ref{Section_2D}; a three-dimensional version would be covered as
well.

Both statements are well known to experts: (i) is the computation of the
\v{C}ech cohomology of a Cantor set, and (ii) is the computation of the
K-theory of the algebra of continuous functions on such a set. We were unable
to find a reference stating them in the form needed here, so we include a
proof for completeness.

\begin{proof}
(i) A compact Hausdorff totally disconnected space is profinite, i.e., it is
the inverse limit $X=\underleftarrow{\lim}X_{i}$ of an inverse system of
finite discrete spaces, \cite[Theorem 1.1.12]{Ribes-Zalesskii}. \v{C}ech
cohomology is continuous with respect to such limits, so $H^{n}\left(
X,\mathbb{Z}\right)  =\underrightarrow{\lim}H^{n}\left(  X_{i},\mathbb{Z}%
\right)  $, see \cite[Chapter X]{Eilenberg-Steenrod}, \cite[Chapter
6]{Spanier}. For a finite discrete space $H^{n}=0$ when $n\geq1$, and $H^{0}$
is the group of all integer-valued functions; passing to the limit gives (i).

(ii) A vector bundle over $X$ is locally trivial, and since $X$ is compact and
totally disconnected, every open cover is refined by a finite partition of $X$
into clopen sets; the bundle is trivial on each member, so it is determined up
to isomorphism by its rank, a locally constant integer function. This gives
$K^{0}\left(  X\right)  \cong C\left(  X,\mathbb{Z}\right)  $. For $K^{1}$,
let $f:X\rightarrow U\left(  n\right)  $ be continuous. By compactness $f$ is
uniformly continuous, so there is a finite clopen partition of $X$ on which
the oscillation of $f$ is smaller than any prescribed $\varepsilon>0$; the
associated locally constant map $g$ satisfies $\left\Vert f-g\right\Vert
<\varepsilon$ and is therefore homotopic to $f$ (for $\varepsilon$ small,
$t\mapsto f\exp\left(  t\log\left(  f^{-1}g\right)  \right)  $ is a homotopy).
Since $g$ takes finitely many values and $U\left(  n\right)  $ is connected,
$g$ is homotopic to a constant. Hence $\left[  X,U\left(  n\right)  \right]
=0$ for every $N$, and $K^{1}\left(  X\right)  =0$. Equivalently: $C\left(
X\right)  $ is an approximately finite-dimensional (AF) algebra when $X$ is a
Cantor set, and $K_{1}$ of an AF algebra vanishes while $K_{0}$ is the group
of locally constant integer functions, \cite[Chapter 7]{Rordam et al},
\cite{Blackadar}.
\end{proof}

For the non-compact Brillouin zones $\mathbb{Q}_{p}^{N}$ and $\left(
\mathbb{Q}_{p}/\mathbb{Z}_{p}\right)  ^{N}$ one uses K-theory with compact
supports, i.e., $K_{j}\left(  C_{0}\left(  \mathrm{BZ}\right)  \right)  $;
since these spaces are increasing unions of compact open subsets of the type
just considered, and K-theory commutes with inductive limits, the same
conclusion holds: $K_{1}=0$, and $K_{0}$ is the group of compactly supported
locally constant integer-valued functions.

Proposition \ref{Prop_nogo} explains, at the level of principle, why the
Jackiw-Rebbi zero mode is not pinned in the $p$-adic self-adjoint model of
Section \ref{Section_sa}. The integer that pins the Archimedean Jackiw-Rebbi
mode to $E=0$ is the jump of a winding number across the wall, that is, a
difference of classes in $K^{1}$ of the Brillouin zone; over $\mathbb{Q}_{p}$
that group is trivial, so the integer does not merely fail to be computable,
it does not exist. The obstruction is the total disconnectedness of the space,
which is exactly the physical hypothesis of $p$-adic quantum mechanics: in
this framework there are no continuous paths in momentum space along which a
phase could wind.

\begin{remark}
Proposition \ref{Prop_nogo} is a statement about invariants extracted from a
band structure, i.e., from a family of matrices indexed by the Brillouin zone.
It does \emph{not} say that $p$-adic models have no topological invariants at
all. The invariants of aperiodic and disordered matter are defined without a
Brillouin zone, as pairings in the K-theory of a noncommutative algebra of
observables; that algebra is not of the form $C\left(  X\right)  $, and its
K-theory need not vanish. This is the subject of the next subsection.
\end{remark}

\subsection{What replaces them}

The aperiodic, disordered and amorphous branches of the theory of topological
materials have developed invariants that do not require a Brillouin zone, and
these are the natural candidates here.

\begin{enumerate}
\item[(1)] \textbf{Noncommutative (real-space) invariants.} Bellissard's
formulation replaces the Brillouin zone by a C*-algebra of observables and
Chern numbers by pairings of K-theory classes with cyclic cocycles,
\cite{Bellissard et al}, \cite{Prodan-Schulz-Baldes}. For $\mathbb{Q}_{p}^{N}$
one has a natural algebra, namely the crossed product of the algebra of
observables by the translation action of $\mathbb{Q}_{p}^{N}$, a canonical
trace given by the Haar measure, and computable K-theory: $\mathbb{Q}_{p}^{N}$
is abelian, hence a-T-menable, so the Baum-Connes conjecture holds for it,
\cite{Higson-Kasparov}. Determining $K_{\ast}$ of this algebra, and which
classes are realized by the Hamiltonians (\ref{Two_Hamiltonians}), is in our
view the central open problem raised by this paper.

\item[(2)] \textbf{Real-space markers on finite levels.} The Bott index,
\cite{Loring-Hastings}, the local Chern marker, \cite{Bianco-Resta}, and the
spectral localizer, \cite{Loring-Schulz-Baldes}, are computable on any finite
set of sites carrying a gapped Hamiltonian, in particular on the finite
quotients $G_{l}=\mathbb{Z}_{p}/p^{l}\mathbb{Z}_{p}$. Whether any of them is
non-zero for a two-dimensional $p$-adic Dirac operator is a well-posed, finite
computation, and it is the concrete form of the question ``are there $p$-adic
Chern insulators?''.

\item[(3)] \textbf{Gap labelling.} For Cantor-like spectra, the values of the
integrated density of states on spectral gaps form a countable group
determined by K-theory, \cite{Bellissard-gap}. For the operators considered
here the spectrum is built on the ladder $p^{\mathbb{Z}}$ and the natural
conjecture is that the integrated density of states takes values in
$\mathbb{Z}[1/p]$, with gap labels given by Haar measures of balls. This seems
to us the most accessible of the problems on this list.

\item[(4)] \textbf{Non-Hermitian invariants.} For $\boldsymbol{H}%
_{\mathrm{nh}}$ the relevant notions are those of non-Hermitian topology:
winding of the complex spectrum, exceptional-point topology, and the non-Bloch
invariants that restore the bulk-boundary correspondence in the presence of
the skin effect, \cite{Yao-Wang}, \cite{Gong et al}, \cite{Bergholtz et al}. A
natural question, with no Archimedean counterpart, is whether an ultrametric
analogue of the skin effect exists: since $\mathbb{Q}_{p}$ has no left and no
right, an accumulation of states would have to occur along the scale
direction, that is, towards small or large balls.
\end{enumerate}

\subsection{\label{Section_positivity}Positivity and the mass profile}

The field $\mathbb{Q}_{p}$ is not ordered, so
the notion of positivity used here, $\pi\left(  x\right)  =1$, is a choice:
any partition $\mathbb{F}_{p}^{\times}=S\sqcup S^{c}$ into two non-empty
classes, extended to $\mathbb{Q}_{p}^{\times}$ through the first digit,
defines a mass profile of the type (\ref{function_m(x)}), and other choices
occur in the literature, see Remark \ref{Remark_positivity} and \cite[Section
6.2.5]{Zuniga-LNM-2016}. The physical content of the model does not depend on
it. The free theory never sees the mass profile; and for any such partition
the profile is locally constant away from the origin, which is all that
Sections \ref{Section_Hamiltonian} and \ref{Section_Compare} use, so that the
chiral symmetry $\Gamma=\sigma_{y}$, the existence of $E=0$ interface states
for $\boldsymbol{H}_{\mathrm{nh}}$ and the quantization condition
(\ref{mass_quantization}) are unaffected, the decay lengths available to a
bound state being radii of balls. Only the normalization constants change,
through the Haar measures of the two regions, which are proportional to $\#S$
and $\#S^{c}$. Whether $\boldsymbol{P}$ exchanges the two regions does depend
on the partition, but this does not reach the Hamiltonian: by Lemma
\ref{Lemma_Gauss}, $\boldsymbol{PHP}^{-1}=-\boldsymbol{H}$ requires $p\equiv3$
$(\operatorname{mod}4)$ under either convention, so the symmetry class of
Section \ref{Section_invariants} is unchanged.

Two variants are of a different nature, and we leave both as open problems.
The first is to let the sign of $x$ depend on $ord(x)$ rather than on its
angular component, declaring, for instance, $x$ to be positive when $ord(x)$
is even: the mass then changes sign with the scale rather than with the
direction, and the interface is a scale rather than a wall. The second is to
replace the character that twists $\widetilde{\boldsymbol{D}}_{x}$, which
changes the operator itself. Self-adjointness forces that character to be
quadratic, by Lemma \ref{Lemma_selfadjoint}, and for $p$ odd the only
non-trivial choices are $\pi$, the unramified character $\varepsilon\left(
x\right)  =\left(  -1\right)  ^{ord\left(  x\right)  }$ and their product,
since $\mathbb{Q}_{p}^{\times}/\left(  \mathbb{Q}_{p}^{\times}\right)  ^{2}$
has order four. For $\varepsilon$ the Kozyrev wavelets remain eigenfunctions,
with eigenvalues $\left(  -1\right)  ^{r-1}p^{1-r}$, so
that the sign of the symbol is governed by the scale, which pairs naturally
with the first variant; the Gauss sum (\ref{Gauss_sum}), the dependence on $p$
$\operatorname{mod}$ $4$ and the closed-form solution of the
interface problem would then have to be reexamined.

\subsection{Open problems}

\begin{enumerate}
\item[(a)] Compute the K-theory of the observable algebra of $p$-adic quantum
mechanics and identify the classes realized by $p$-adic Dirac operators.

\item[(b)] Prove or disprove that the integrated density of states of
$\boldsymbol{H}_{\mathrm{sa}}$ takes values in $\mathbb{Z}[1/p]$.

\item[(c)] Decide, by computing a Bott index or a spectral localizer on
$G_{l}$, whether a two-dimensional $p$-adic Dirac operator can be
topologically non-trivial in a real-space sense, despite Proposition
\ref{Prop_nogo}.

\item[(d)] Decide whether $\boldsymbol{H}_{\mathrm{sa}}$ with the
sign-changing mass (\ref{function_m(x)}) has bound states inside the gap
(Remark \ref{Remark_origin}), and determine their energies.

\item[(e)] Define a velocity, and a current, for the two-dimensional interface
states of Section \ref{Section_2D}.

\item[(f)] Extend the analysis to multiplicative characters of order greater
than two, where the natural invariants may be torsion-valued rather than integer-valued.

\item[(g)] Decide whether the stochastic-matrix property of \cite{Zuniga-CTQW}
survives for the walks of Theorem \ref{Prop_CTQW}, and whether the interface
states survive the discretization; this is the sixth open problem of
\cite{Zuniga-CTQW}.
\end{enumerate}

\section{Conclusions}

We have introduced $p$-adic versions of the 1D and 2D Jackiw-Rebbi model and
analyzed them in two inequivalent implementations. The self-adjoint model,
built from the dictionary $-i\partial_{x}\leftrightarrow\widetilde
{\boldsymbol{D}}_{x}$, is gapped when the mass is constant and has the chiral
symmetry $\Gamma=\sigma_{y}$; the non-Hermitian model, obtained by keeping the
factor $-i$, supports interface states at exactly $E=0$, but only when the
masses satisfy the quantization condition $m_{1,2}v\in p^{\mathbb{Z}}$, and
these states are second-order exceptional points of an operator with a
$\mathcal{PT}$-type symmetry.

The physical content of the construction is the subject of Section
\ref{Section_physics}. The free $p$-adic Dirac Hamiltonian is diagonalized by
a basis of compactly supported wavelets; it therefore has pure point spectrum
and a complete family of strictly localized stationary states, neither of
which exists in the Archimedean theory, where the free Dirac operator has
purely absolutely continuous spectrum and no eigenvalues. Momentum takes its
values on a geometric ladder, so the two theories agree at energies well below
the mass gap and separate above it by a fixed factor $p$ between consecutive
levels. At an interface, the geometry quantizes the masses that admit a
zero-energy state. The dynamics, moreover, is the limit of continuous-time
quantum walks on finite hierarchical graphs, in the sense of
\cite{Zuniga-CTQW}, so that it is prescribed in implementable terms at every
finite level of resolution. These three statements are our answer, for this
model, to the question of whether $p$-adic quantum mechanics has physical
content: it reproduces the physical picture of the Archimedean theory by
mechanisms that have no Archimedean counterpart.

We have also shown that the symmetry class of these operators is controlled by
$p$ $\operatorname{mod}$ $4$, through the sign of the quadratic Gauss sum, and
that Chern and winding numbers cannot be defined from a $p$-adic Brillouin
zone, which is totally disconnected. The invariants that survive are the
real-space, K-theoretic ones developed for aperiodic and disordered matter.
The Jackiw-Rebbi mechanism therefore survives the passage to an ultrametric
space, while the topological protection of its zero mode does not.

\section*{Acknowledgements}

During the preparation of this work, the author used Claude (Anthropic; model
claude-opus-5) to help him with three different tasks. First, to improve the
clarity and readability of select sections. Second, to carry out the
calculations of the symmetries (Lemma \ref{Lemma_AZ}) and the counting of the
elementary solutions and of the chiral zero modes; to the best of the author's
knowledge, these calculations are correct. Third, to develop Figure
\ref{Fig_dispersion}. After using this tool, the author reviewed and edited the
content as needed and takes full responsibility for the content of the
publication.


\begin{thebibliography}{99}                                                                                               %


\bibitem {Capellmann}Capellmann H. Space-Time in Quantum Theory. Found Phys
51 (2021), no. 3, Paper No. 44.

\bibitem {Amelino-Camelia}Amelino-Camelia G. Quantum-Spacetime Phenomenology.
Living Rev. Relativ. 16, 5 (2013).

\bibitem {Bronstein}Bronstein M. Republication of: Quantum theory of weak
gravitational fields. Gen Relativ Gravit 44, 267--283 (2012).

\bibitem {Rovelli}Rovelli C, Vidotto F. Covariant Loop Quantum Gravity: An
Elementary Introduction to Quantum Gravity and Spinfoam Theory. Cambridge
University Press; 2014.

\bibitem {Volovich}Volovich I. V. Number theory as the ultimate physical
theory. $p$-Adic Numbers Ultrametric Anal. Appl. 2 (2010), no. 1, 77--87.

\bibitem {Varadarajan}Varadarajan V. S. Reflections on quanta, symmetries, and
supersymmetries. Springer, New York, 2011.

\bibitem {V-V-Z}Vladimirov V. S., Volovich I. V., Zelenov E. I. $p$-Adic
analysis and mathematical physics. World Scientific, 1994.

\bibitem {Dragovivh et al}Dragovich B., Khrennikov A. Y., Kozyrev S. V.,
Volovich, I. V. On $p$-adic mathematical physics. $p$-Adic Numbers Ultrametric
Anal. Appl. 1 (2009), no. 1, 1--17.

\bibitem {Kozyrev}Kozyrev S. V. Wavelet theory as $p$-adic spectral analysis.
Izv. Math. 66 (2002), no. 2, 367--376.

\bibitem {Zuniga-Textbook}Z\'{u}\~{n}iga-Galindo W. A. $p$-Adic Analysis:
Stochastic Processes and Pseudo-Differential Equations. De Gruyter, 2025.

\bibitem {Serre}Serre Jean-Pierre. Lie algebras and Lie groups. W. A.
Benjamin, Inc., New York-Amsterdam, 1965.

\bibitem {V-V-QM}Vladimirov V. S., Volovich I. V. $p$-Adic quantum mechanics.
Comm. Math. Phys. 123 (1989), no. 4, 659--676.

\bibitem {Varadarajan-2001}Varadarajan V. S. Non-Archimedean models for
space-time. Mod. Phys. Lett. A 16 (2001), 387--395.

\bibitem {Vourdas}Vourdas A. Quantum systems with finite Hilbert space. Rep.
Progr. Phys. 67 (2004), no. 3, 267--320.

\bibitem {Anashin}Anashin V. Causality: the $p$-adic theory. Springer, Cham,
2025.

\bibitem {Aniello et al}Aniello P., Mancini S., Parisi V. A $p$-adic model of
quantum states and the $p$-adic qubit. Entropy 25 (2023), no. 1, 86.

\bibitem {Zelenov}Zelenov E. I. $p$-Adic model of quantum mechanics and
quantum channels. Proc. Steklov Inst. Math. 285 (2014), no. 1, 132--144.

\bibitem {Urban}Urban Roman. A note on $p$-adic Schr\"{o}dinger equations
with nonlocal potentials. J. Math. Phys. Anal. Geom. 22 (2026), no. 4,
385--398.

\bibitem {Zuniga-QM-2}Z\'{u}\~{n}iga-Galindo W. A. $2$-Adic quantum mechanics,
continuous-time quantum walks, and the space discreteness. Fortschr. Phys.
2025, Paper No. e70019.

\bibitem {Zuniga-AP}Z\'{u}\~{n}iga-Galindo W. A. The $p$-Adic Schr\"{o}dinger
equation and the two-slit experiment in quantum mechanics. Ann. Physics 469
(2024), Paper No. 169747.

\bibitem {Zuniga-PhA}Z\'{u}\~{n}iga-Galindo W. A. $p$-Adic quantum mechanics,
the Dirac equation, and the violation of Einstein causality. J. Phys. A 57
(2024), no. 30, Paper No. 305301, 29 pp.

\bibitem {Zuniga-Mayes}Z\'{u}\~{n}iga-Galindo W. A., Mayes Nathaniel P.
$p$-Adic quantum mechanics, infinite potential wells, and continuous-time
quantum walks. Rev. Math. Phys. 38 (2026), no. 6, Paper No. 2550027, 32 pp.

\bibitem {Zuniga-Chacon}Z\'{u}\~{n}iga-Galindo W. A., Chac\'{o}n-Cort\'{e}s L.
F. Continuous-time quantum Markov chains and discretizations of $p$-adic
Schr\"{o}dinger equations: comparisons and simulations. J. Stat. Mech. Theory
Exp. 2026, no. 9, Paper No. 093401, 55 pp.

\bibitem {Zuniga-CTQW}Z\'{u}\~{n}iga-Galindo W. A. $p$-Adic Dirac equations,
continuous-time quantum walks, and quantum networks. arXiv:2609.04358.

\bibitem {Mulkne-Blumen}M\"{u}lken O., Blumen A. Continuous-time quantum
walks: models for coherent transport on complex networks. Phys. Rep. 502
(2011), no. 2-3, 37--87.

\bibitem {Venegas-Andraca}Venegas-Andraca Salvador El\'{\i}as. Quantum walks:
a comprehensive review. Quantum Inf. Process. 11 (2012), no. 5, 1015--1106.

\bibitem {Jackiw-Rebbi}Jackiw R., Rebbi C. Solitons with fermion number 1/2.
Phys. Rev. D 13 (1976), no. 12, 3398.

\bibitem {Goldstone}Goldstone J., Wilczek F. Fractional quantum numbers on
solitons. Phys. Rev. Lett. 47 (1981), 986.

\bibitem {Su et al}Su W. P., Schrieffer J. R., Heeger A. J. Solitons in
polyacetylene. Phys. Rev. Lett. 42 (1979), 1698.

\bibitem {S-Q-Shen-Book}Shen SQ. Topological insulators: Dirac equation in
condensed matters. Berlin; New York: Springer, 2012.

\bibitem {Kotetes}Kotetes P. (2019). Topological Insulators (IOP Concise
Physics). Morgan \& Claypool Publishers, 2019.

\bibitem {Thaller}Thaller Bernd. The Dirac equation. Texts Monogr. Phys.
Springer-Verlag, Berlin, 1992.

\bibitem {Taibleson}Taibleson M. H. Fourier analysis on local fields.
Princeton University Press, 1975.

\bibitem {Alberio et al}Albeverio S., Khrennikov A. Y., Shelkovich V. M.
Theory of $p$-adic distributions: linear and nonlinear models. London
Mathematical Society Lecture Note Series, 370. Cambridge University Press, 2010.

\bibitem {Halmos}Halmos P. Measure Theory. D. Van Nostrand Company Inc., New
York, 1950.

\bibitem {Ireland-Rosen}Ireland K., Rosen M. A classical introduction to
modern number theory. Second edition. Graduate Texts in Mathematics 84,
Springer-Verlag, New York, 1990.

\bibitem {Zuniga-LNM-2016}Z\'{u}\~{n}iga-Galindo W. A. Pseudodifferential
equations over non-Archimedean spaces. Lectures Notes in Mathematics 2174,
Springer, 2016.

\bibitem {Bender-Boettcher}Bender C. M., Boettcher S. Real spectra in
non-Hermitian Hamiltonians having $\mathcal{PT}$ symmetry. Phys. Rev. Lett. 80
(1998), 5243--5246.

\bibitem {Heiss}Heiss W. D. The physics of exceptional points. J. Phys. A 45
(2012), 444016.

\bibitem {Ashida et al}Ashida Y., Gong Z., Ueda M. Non-Hermitian physics. Adv.
Phys. 69 (2020), no. 3, 249--435.

\bibitem {Bergholtz et al}Bergholtz E. J., Budich J. C., Kunst F. K.
Exceptional topology of non-Hermitian systems. Rev. Mod. Phys. 93 (2021), 015005.

\bibitem {Gong et al}Gong Z., Ashida Y., Kawabata K., Takasan K., Higashikawa
S., Ueda M. Topological phases of non-Hermitian systems. Phys. Rev. X 8
(2018), 031079.

\bibitem {Yao-Wang}Yao S., Wang Z. Edge states and topological invariants of
non-Hermitian systems. Phys. Rev. Lett. 121 (2018), 086803.

\bibitem {Newton-Wigner}Newton T. D., Wigner E. P. Localized states for
elementary systems. Rev. Mod. Phys. 21 (1949), no. 3, 400.

\bibitem {Felimng et al}Fleming G. N. Nonlocal properties of stable
particles. Phys. Rev. 139 (1965), no. 4B, B963.

\bibitem {Hegerfeldt}Hegerfeldt G. C. Remark on causality and particle
localization. Phys. Rev. D 10 (1974), no. 10, 3320.

\bibitem {Hegerfeldt et al}Hegerfeldt G. C., Ruijsenaars S. N. M. Remarks on
causality, localization, and spreading of wave packets. Phys. Rev. D 22
(1980), no. 2, 377.

\bibitem {Eberhard et al}Eberhard P. H., Ross R. R. Quantum field theory
cannot provide faster-than-light communication. Found. Phys. Lett. 2 (1989),
no. 2, 127--149.

\bibitem {Zuniga-Galindo-ultimo}Z\'{u}\~{n}iga-Galindo W. A. Quantum
mechanics, non-locality, and the space discreteness hypothesis. Ann. Physics
489 (2026), Paper No. 170459, 17 pp.

\bibitem {Reed-Simon}Reed M., Simon B. Methods of Modern Mathematical Physics
I: Functional Analysis. Academic Press, New York, 1980.

\bibitem {Kitaev-table}Kitaev A. Periodic table for topological insulators and
superconductors. AIP Conf. Proc. 1134 (2009), 22--30.

\bibitem {Chiu et al}Chiu C.-K., Teo J. C. Y., Schnyder A. P., Ryu S.
Classification of topological quantum matter with symmetries. Rev. Mod. Phys.
88 (2016), 035005.

\bibitem {Rudin-groups}Rudin W. Fourier Analysis on Groups. Interscience
Publishers, New York, 1962.

\bibitem {Hewitt-Ross}Hewitt E., Ross K. A. Abstract Harmonic Analysis I.
Springer, Berlin, 1963.

\bibitem {Prodan-Schulz-Baldes}Prodan E., Schulz-Baldes H. Bulk and Boundary
Invariants for Complex Topological Insulators: From K-Theory to Physics.
Springer, 2016.

\bibitem {Eilenberg-Steenrod}Eilenberg S., Steenrod N. Foundations of
Algebraic Topology. Princeton University Press, 1952.

\bibitem {Spanier}Spanier E. H. Algebraic Topology. McGraw-Hill, New York, 1966.

\bibitem {Atiyah}Atiyah M. F. K-Theory. W. A. Benjamin, New York, 1967.

\bibitem {Blackadar}Blackadar B. K-Theory for Operator Algebras. 2nd ed.,
Cambridge University Press, 1998.

\bibitem {Rordam et al}R{\o }rdam M., Larsen F., Laustsen N. J. An
Introduction to K-Theory for C*-Algebras. Cambridge University Press, 2000.

\bibitem {Ribes-Zalesskii}Ribes L., Zalesskii P. Profinite Groups. 2nd ed.,
Springer, Berlin, 2010.

\bibitem {Bellissard et al}Bellissard J., van Elst A., Schulz-Baldes H. The
noncommutative geometry of the quantum Hall effect. J. Math. Phys. 35 (1994), 5373--5451.

\bibitem {Higson-Kasparov}Higson N., Kasparov G. E-theory and KK-theory for
groups which act properly and isometrically on Hilbert space. Invent. Math.
144 (2001), 23--74.

\bibitem {Loring-Hastings}Loring T. A., Hastings M. B. Disordered topological
insulators via C*-algebras. EPL 92 (2010), 67004.

\bibitem {Bianco-Resta}Bianco R., Resta R. Mapping topological order in
coordinate space. Phys. Rev. B 84 (2011), 241106(R).

\bibitem {Loring-Schulz-Baldes}Loring T. A., Schulz-Baldes H. The spectral
localizer for even index pairings. J. Noncommut. Geom. 14 (2020), 1--23.

\bibitem {Bellissard-gap}Bellissard J. Gap labelling theorems for
Schr\"{o}dinger operators. In: From Number Theory to Physics, Springer, 1992, 538--630.

\bibitem {Kochubei}Kochubei A.N. Pseudo-differential equations and stochastics
over non-Archimedean fields. Marcel Dekker, New York, 2001.

\end{thebibliography}
\end{document}